\documentclass[aps,prc,amsfonts,preprintnumbers,superscriptaddress,showpacs,nofootinbib]{revtex4}

\usepackage{graphics}
\usepackage{psfrag}
\usepackage{epsfig}
\usepackage{epsf}
\usepackage{float}
\usepackage{bbm}

\newcommand{\beq}{\begin{equation}}
\newcommand{\eeq}{\end{equation}}
\newcommand{\beqa}{\begin{eqnarray}}
\newcommand{\eeqa}{\end{eqnarray}}
\newcommand{\nn}{\nonumber \\ }

\newcommand{\II}[4]{I^{\left(#1\right)}_{\left(#2,#3,#4 \right)}}
\newcommand{\IIR}[4]{I_{\rm red \ \left(#2,#3,#4 \right)}^{\left( #1 \right)}}

\newcommand{\vek}[4]{\left(\vec{#1}_{#3}\cdot\vec{#2}_{#4} \right)}
\newcommand{\xp}[4]{\left[\vec{#1}_{#3}\times \vec{#2}_{#4} \right]^3}
\newcommand{\cp}[4]{\left[\vec{#1}_{#3}\times \vec{#2}_{#4} \right]}
\newcommand{\spr}[6]{\left(\vec{#1}_{#4} \cdot\left[ \vec{#2}_{#5}\times\vec{#3}_{#6}\right] \right)}
\newcommand{\one}{\mathbbm{1}}

\begin{document}

\preprint{FZJ-IKP-TH-2009-23}
\preprint{HISKP-TH-09/21}

\title{Two-pion exchange electromagnetic current in chiral effective field
  theory using
  the method of unitary transformation}

\author{S.~K\"olling}
\email[]{Email: s.koelling@fz-juelich.de}
\affiliation{Forschungszentrum J\"ulich, Institut f\"ur Kernphysik (IKP-3) and
J\"ulich Center for Hadron Physics, \\
             D-52425 J\"ulich, Germany}
\affiliation{Helmholtz-Institut f\"ur Strahlen- und Kernphysik (Theorie)
and Bethe Center for Theoretical Physics,\\
 Universit\"at Bonn, D-53115 Bonn, Germany}
\author{E.~Epelbaum}
\email[]{Email: e.epelbaum@fz-juelich.de}
\affiliation{Forschungszentrum J\"ulich, Institut f\"ur Kernphysik (IKP-3) and
J\"ulich Center for Hadron Physics, \\
             D-52425 J\"ulich, Germany}
\affiliation{Helmholtz-Institut f\"ur Strahlen- und Kernphysik (Theorie)
and Bethe Center for Theoretical Physics,\\
 Universit\"at Bonn, D-53115 Bonn, Germany}
\author{H.~Krebs}
\email[]{Email: hkrebs@itkp.uni-bonn.de}
\affiliation{Helmholtz-Institut f\"ur Strahlen- und Kernphysik (Theorie)
and Bethe Center for Theoretical Physics,\\
 Universit\"at Bonn, D-53115 Bonn, Germany}
\author{U.-G.~Mei{\ss}ner}
\email[]{Email: meissner@itkp.uni-bonn.de}
\affiliation{Helmholtz-Institut f\"ur Strahlen- und Kernphysik (Theorie)
and Bethe Center for Theoretical Physics,\\
 Universit\"at Bonn, D-53115 Bonn, Germany}
\affiliation{Forschungszentrum J\"ulich, Institut f\"ur Kernphysik (IKP-3) and
J\"ulich Center for Hadron Physics, \\
             D-52425 J\"ulich, Germany}
\affiliation{Forschungszentrum J\"ulich, Institut for Advanced Simulation,
 D-52425 J\"ulich, Germany}
\date{\today}

\begin{abstract}
We derive the leading two-pion exchange contributions to the two-nucleon
electromagnetic current operator in the framework of chiral effective field
theory using the method of unitary transformation. Explicit results for the
current  and charge densities are given in
momentum and coordinate space. 

\end{abstract}


\maketitle

%\vspace{-0.2cm}

%%%%%%%%%%%%%%%%%%%%%%%%%%%%%%%%%%%%%%%%%%%%%%%%%%%%%%%%%%%%%%%%%%%%%%%%%%%%%%%%%
\section{Introduction}
\def\theequation{\arabic{section}.\arabic{equation}}
\setcounter{equation}{0}
\label{sec:intro}

Chiral effective field theory (EFT) provides a systematic and model-independent
framework to analyze low-energy hadronic processes in harmony with the
spontaneously broken approximate chiral symmetry of QCD. This approach has
been successfully  applied to the derivation of the nuclear
forces and, more recently, also  hyperon-nucleon and hyperon-hyperon
interactions, see Refs.~\cite{Bedaque:2002mn,Epelbaum:2005pn,Epelbaum:2008ga}
for review articles. Exchange vector and axial currents in nuclei have
also been studied in the framework of chiral EFT. In their pioneering 
work, Park, Min and Rho applied heavy-baryon chiral perturbation theory to
derive exchange axial \cite{Park:1993jf} and vector \cite{Park:1995pn}
currents at the one-loop level for small values of the photon momentum
focusing, in particular, on the axial-charge and magnetic moment operators,
respectively. These calculations were carried out employing time-ordered
perturbation theory to extract non-iterative
contributions to the amplitude. The resulting exchange vector currents
were applied within a hybrid approach to analyze magnetic moments and 
radiative capture cross sections of thermal neutrons on light nuclei 
\cite{Park:1995pn,Song:2007bj,Lazauskas:2009nw,Song:2008zf} as well as some polarization
observables in  radiative neutron capture on the proton
\cite{Park:1999sz}. For applications  to various
electroweak few-nucleon reactions of astrophysical interest see
\cite{Park:1998wq,Park:2001jn,Ando:2001es,Park:2002yp,Kubodera:2004zm}.  

Deuteron electromagnetic properties
\cite{Phillips:1999am,Walzl:2001vb,Phillips:2003jz,Phillips:2006im,Valderrama:2007ja},
Compton scattering on 
the deuteron \cite{Beane:2004ra,Choudhury:2004yz} and, more 
recently, on $^3$He \cite{Shukla:2008zc} as well as pion electro- and
photoproduction and the corresponding capture reactions 
\cite{Beane:1997iv,Bernard:1999ff,Krebs:2002qr,Krebs:2004ir,Gardestig:2005pp,
  Lensky:2005hb}  have also been addressed in the framework
of chiral EFT. However, to the best of our knowledge, no
applications have so far been performed to electron and photon inelastic  
few-nucleon reactions with the momentum transfer of the order
$M_\pi$ where a lot of experimental data are available, see \cite{Golak:2005iy} for a
recent review article on the theoretical achievements in this field based on
conventional framework. Recent progress in the accurate description of the
two- \cite{Epelbaum:2004fk,Entem:2003ft} and more-nucleon systems \cite{Epelbaum:2002vt}
within the chiral EFT, see also \cite{Epelbaum:2008ga} and references therein, gives a 
strong motivation to apply this framework to the abovementioned
processes. This requires the knowledge of the consistent electromagnetic
exchange current operator for non-vanishing values of the photon momentum. While the
leading two-nucleon  contributions to the exchange current arise from one-pion
exchange and are well known, the corrections at the one-loop level have not
yet been completely worked out. An important step in this direction was done
recently by Pastore et al.~\cite{Pastore:2008ui,Pastore:2009is} who considered
the electromagnetic two-body 
current density at the leading one-loop order based on time-ordered
perturbation theory. In the present work, we calculate the leading two-pion
exchange two-nucleon four-current operator in chiral EFT based on the method of unitary
transformation which we used to derive nuclear forces in
Refs.~\cite{Epelbaum:1998ka,Epelbaum:1999dj,Epelbaum:2002gb,
  Epelbaum:2006eu,Epelbaum:2007us}. Our work provides an important check
of the results presented in Refs.~\cite{Pastore:2008ui,Pastore:2009is}
but also differs from these works in several important respects. First, as
already pointed out, we use a completely different method to compute the
current operator. Secondly, we also give results for the exchange charge
density which, to the best of our knowledge, have not yet been published
before. Finally, we evaluate analytically all loop integrals to obtain  
a representation in momentum space in terms of the standard loop functions and
the three-point functions. The latter are reduced to a form which can be
easily treated numerically. We also succeeded to analytically carry out the
Fourier transformation for all contributions leading to an extremely compact
representation of the current and charge densities in coordinate space.
Notice  that contrary to \cite{Pastore:2008ui,Pastore:2009is}, we do
not treat the $\Delta$(1232) isobars as explicit degrees of freedom in this
work.   

Our manuscript is organized as follows. In section \ref{sec1} we provide a short
summary of the method of unitary transformation, explain the adopted power
counting scheme, list all relevant terms in the effective chiral Lagrangian
and present our results for the exchange current and charge densities
in momentum space. The expressions in coordinate space are given in section
\ref{sec2}. The results of our work are summarized in section
\ref{sec:summary}. The formal operator structure of the effective
electromagnetic current can be found in appendix \ref{app1}, while appendix
\ref{app2} contains the complete momentum-space representation of the
current. Evaluation of the three-point functions entering this representation
is detailed in appendix \ref{app3}. In appendix \ref{app4} we verify that the
obtained expressions for the current fulfill the continuity equation. Finally,
appendix \ref{app5} collects the formulae needed to obtain the
coordinate-space representation given in section \ref{sec2}.

%%%%%%%%%%%%%%%%%%%%%%%%%%%%%%%%%%%%%%%%%%%%%%%%%%%%%%%%%%%%%%%%%%%%%%%%%%%%%%%%%
\section{Nuclear currents using the method of unitary transformation}
\def\theequation{\arabic{section}.\arabic{equation}}
\setcounter{equation}{0}
\label{sec1}

We begin with a brief reminder about the method of unitary
transformation. Consider the time-independent Schr\"odinger equation 
for interacting pions and nucleons in the absence of electromagnetic sources
\beq
\label{schroed1}
(H_0 + H_I) | \Psi \rangle = E | \Psi \rangle\,,
\eeq
where $|\Psi \rangle$ denotes an eigenstate of the Hamiltonian $H$ 
with the eigenvalue $E$. Let $\eta$ ($\lambda$)  be 
projection operators onto the purely nucleonic (the remaining) part of the
Fock space satisfying $\eta^2 = \eta$, $\lambda^2 = \lambda$, $ \eta \lambda 
= \lambda \eta = 0$ and $\lambda + \eta = {\bf 1}$. 
To describe the dynamics of few- and many-nucleon systems below the pion
production threshold it is advantageous to  
project Eq.~(\ref{schroed1}) onto the $\eta$-subspace of the 
full Fock space. The resulting effective equation can then be solved using 
the standard methods of few- or many-body physics. The decoupling of the
$\eta$- and $\lambda$-subspaces can be achieved via
an appropriately chosen unitary transformation \cite{Okubo:1954aa,Fukuda:1954aa}
\beq
\tilde H \equiv U^\dagger H U = \left( \begin{array}{cc} \eta \tilde H \eta  &
    0 
\\ 0 & \lambda \tilde H \lambda \end{array} \right)\,,
\eeq
see Ref.~\cite{Krebs:2004st} for an alternative approach. 
Following Okubo \cite{Okubo:1954aa}, the unitary operator $U$ can be parametrized as
\begin{equation}
\label{5.9}
U = \left( \begin{array}{cc} \eta (1 +  A^\dagger  A )^{- 1/2} & - 
 A^\dagger ( 1 +  A A^\dagger )^{- 1/2} \\
 A ( 1 +  A^\dagger  A )^{- 1/2} & 
\lambda (1 +  A  A^\dagger )^{- 1/2} \end{array} \right)~,
\end{equation}
in terms of the operator $A= \lambda  A \eta$ which has to satisfy 
the decoupling equation 
\begin{equation}
\label{5.10}
\lambda \left( H - \left[ A, \; H \right] - A H A \right) \eta = 0
\end{equation}
in order for the transformed Hamiltonian $\tilde H$ to be of  block-diagonal form. 
The effective $\eta$-space potential $V$ can be expressed in terms of the
operator $A$ via:  
\beq
\label{effpot}
{V} =  \eta (\tilde H  - H_0 )  = \eta \bigg[ (1 + A^\dagger A)^{-1/2} (H +
A^\dagger H + H A + A^\dagger H A )   
(1 + A^\dagger A)^{-1/2} - H_0 \bigg] \eta~.
\eeq

The unitary transformation $U$ and the effective potential $V$ can be
calculated perturbatively based on the most general effective chiral
Lagrangian utilizing the chiral power counting. In Ref.~\cite{Epelbaum:2007us}, a convenient
formulation of the power counting has been presented. The
low-momentum dimension $\nu$ of the effective potential, $V$, $V \sim
\mathcal{O}(Q/\Lambda)^\nu$ with $Q$ and $\Lambda$ refering to the soft and
hard scales of the order of the pion and $\rho$-meson masses, respectively, is
given (modulo the normalization constant $-2$) by the overall inverse mass dimension
of the coupling constants entering the expression for $V$: 
\beq
\label{pow_fin}
\nu = -2 + \sum V_i \kappa_i \,, \quad \quad \kappa_i = d_i + \frac{3}{2} n_i + p_i - 4\,.
\eeq
Here, where $V_i$ is the number of vertices of type $i$ while $d_i$, $n_i$ and
$p_i$ refer to the number of derivatives or $M_\pi$-insertions, nucleon field
and pion field operators, respectively. Further, $\kappa_i$ is simply the 
canonical field dimension of a vertex of type $i$ (up to the additional
constant $-4$). Writing the effective chiral Hamiltonian $H$ as 
\begin{equation}
\label{n11}
H = \sum_{\kappa = 1}^{\infty} H^{(\kappa )} \,,
\end{equation}
the operator $A$ can be calculated recursively,
\beq
\label{n13}
A = \sum_{\alpha = 1}^\infty A^{(\alpha )}\,, \quad 
A^{( \alpha )} = \frac{1}{E_\eta - E_\lambda} \lambda \bigg[ H^{(\alpha )} + \sum_{i =
    1}^{\alpha -1} H^{(i)} A^{(\alpha -i)} - \sum_{i=1}^{\alpha -1} A^{(\alpha -i)} H^{(i)}
- \sum_{i = 1}^{\alpha -2} \; \sum_{j =1}^{\alpha - j - 1} A^{(i)} H^{(j)} A^{(\alpha -i-j)} 
\bigg] \eta
\,. 
\eeq
Here, $E_\eta$ ($E_\lambda$) refers to the free energy of nucleons (nucleons and pions)
in the state $\eta$ ($\lambda$). The expressions for the unitary operator and
the effective potential then  
follow immediately by substituting Eqs.~(\ref{n11}) and (\ref{n13}) into
Eq.~(\ref{effpot}). It is important to emphasize that Eq.~(\ref{5.9}) does not
provide the most 
general parametrization of the unitary operator. Moreover, as found in
Ref.~\cite{Epelbaum:2007us}, the subleading contributions to the
three-nucleon force obtained using the parametrization in
Eq.~(\ref{5.9}) cannot be renormalized. To restore renormalizability at the
level of the Hamilton
operator additional unitary transformation $U'$ in the $\eta$-subspace of the Fock
space had to be employed, $\eta U ' \eta {U '}^\dagger \eta =  \eta {U
  '}^\dagger \eta U ' \eta  = \eta$, whose explicit form at lowest non-trivial
order is given in that work. 

It is, in principle, straightforward to extend this formalism to low-energy electromagnetic
reactions such as e.g.~electron scattering off light nuclei, see
\cite{Gari:1976kj,Hyuga:1977cj,Eden:1995rf} for some
early applications of the method of unitary transformation to the
derivation of the exchange currents. Here and in what
follows, we restrict ourselves to the one-photon-exchange approximation to the
scattering amplitude. The effective nuclear current operator $\eta J^\mu_{\rm eff}
( x ) \eta$ acting in the $\eta$-space is then defined according to
\beq
\label{current_def}
\langle \Psi_f | J^\mu ( x ) | \Psi_i \rangle =  \langle \phi_f | \eta {U
  '}^\dagger \eta U^\dagger
J^\mu ( x ) U \eta U ' \eta | \phi_i \rangle \equiv  \langle \phi_f | 
\eta J^\mu_{\rm eff} ( x )  \eta | \phi_i \rangle \,,
\eeq
where  $\eta |\phi_{i,f} \rangle = \eta {U'}^\dagger \eta U^\dagger
|\Psi_{i,f} \rangle$ denote the transformed states and we have
omitted the components $\lambda |\phi_{i,f} \rangle$ which is justified as
long as one stays below the pion production threshold. In the above
expression, $J^\mu (x)$ denotes the hadronic current density which enters the
effective Lagrangian $\mathcal{L}_{\pi N
  \gamma}$ describing the interaction of pions and nucleons with an
external electromagnetic field $\mathcal{A}^\mu$ and is given by 
\beq
J^\mu (x) = \partial_\nu \frac{\partial \mathcal{L}_{\pi N \gamma}}{\partial
  (\partial_\nu \mathcal{A}_\mu)} - \frac{\partial \mathcal{L}_{\pi N \gamma}}{\partial
  \mathcal{A}_\mu}\,.
\eeq
Notice that contrary to the Hamilton operator, the unitarily transformed
current does, in general, not have the
block-diagonal form, i.e.~$\eta U^\dagger J^\mu ( x ) U \lambda \neq 0$.  
Again, it is important to realize that the above definition of $\eta J^\mu_{\rm eff}
( x ) \eta$ does not fully incorporate the freedom in the choice of unitary
transformations. Thus, one might expect that this formulation yields 
the effective current operator which is not renormalizable by a redefinition of
the low-energy constants (LECs) entering the underlying Lagrangian. Indeed,
renormalizability of the 
effective current operator implies highly non-trivial constraints in the case
of one-pion exchange contributions at the one-loop level since all $\beta$-functions of
the LECs $l_i$ from
$\mathcal{L}_{\pi}$ \cite{Gasser:1984ux,Ecker:1995rk} and 
$d_i$ from  $\mathcal{L}_{\pi N}$
\cite{Fettes:1998ud,Gasser:2002am} are fixed. We have
verified that the ultraviolet divergences 
entering the expressions for the one-pion exchange contributions using the
formulation based on the $\mathcal{A}^\mu$-independent unitary transformation as
described above can indeed not be completely removed by the redefinition of the
corresponding LECs. Thus, a more general parametrization of the unitary
transformation is required in order to restore renormalizability of the nuclear current. 
This can be achieved if one allows for the unitary operator to depend explicitly
on the external electromagnetic field, $U ( \mathcal{A}^\mu )$.  
%Clearly, to the Hamilton
%operator $H_{\pi N \gamma}$ and keeping only terms linear in $\mathcal{A}^\mu$. 
The operator $U ( \mathcal{A}^\mu )$ then has to be chosen in such a way that the transformed
Hamiltonian $U^\dagger ( \mathcal{A}^\mu ) H_{\pi N \gamma} U (
\mathcal{A}^\mu )$ is block-diagonal (with 
respect to the $\eta$- and $\lambda$-spaces) and coincides with the one given
in Ref.~\cite{Epelbaum:2007us} when the external electromagnetic field is switched off. 
The effective nuclear current operator $\eta J^\mu_{\rm eff} ( x ) \eta$
in this more general formulation receives additional contributions which are
not included in Eq.~(\ref{current_def}) and result from $\mathcal{A}^\mu$-dependent
pieces of $U ( \mathcal{A}^\mu )$ in the expression $U^\dagger ( \mathcal{A}^\mu ) H_{\pi N
  \gamma} U ( \mathcal{A}^\mu )$ whose form is determined by renormalizability of the
resulting nuclear current operator.  These additional
terms in  $\eta J^\mu_{\rm eff} ( x ) \eta$ are found to have no effect
on the two-pion exchange current and will be discussed in detail
in a separate publication \cite{Koelling:future} devoted to the one-pion exchange
contributions.  Finally, we emphasize that the power counting employed in the
present work implies the following restrictions on the photon momentum $k$ in the
two-nucleon rest frame
\beq
| \vec k | \sim \mathcal{O}\left( M_\pi \right)\,, \quad \quad 
k^0  \sim \mathcal{O}\left( \frac{M_\pi^2}{m} \right) \ll M_\pi\,,
\eeq 
where $M_\pi$ and $m$ refer to the pion and nucleon masses,
respectively. For the kinematics with $k^0  \sim \mathcal{O}\left( M_\pi
\right)$, one will have to systematically keep track of the new soft momentum
scale $\sqrt{M_\pi  m}$. This goes beyond the scope of the present work. 

For the calculation of the leading two-pion exchange two-nuclear current
operator in the present work we only need the leading pion and pion-nucleon
terms in the effective Lagrangian
\beqa
\label{lagrange}
  \mathcal{L}_{\pi\pi}^{(2)}&=&\frac{F_\pi^2}{4} \, \mbox{tr}
  \left[  D_\mu U D^\mu U^\dagger 
+ M_\pi^2 ( U + U^\dagger ) \right]\,, \nn
  \mathcal{L}_{\pi N}^{(1)} &=& N^\dagger(i \,v \cdot D + g_A \, S\cdot u
  \,)N \,,
\eeqa
where the superscript $i$ in $\mathcal{L}^{(i)}$ denotes the number of
derivatives and/or pion mass insertions. Here, $F_\pi$ ($g_A$) is the 
pion decay constant (the nucleon axial-vector coupling), 
$N$ represents a nucleon field in the heavy-baryon formulation and 
$S_\mu=\frac{1}{2}\gamma_5 \sigma_{\mu\nu}v^\nu$ is the Pauli-Lubanski spin
vector which reduces to 
$S^\mu=(0,\frac{1}{2}\vec{\sigma})$ for $v_\mu=(1,0,0,0)$. At the order we are
working and for the contributions to the current operator considered in the
present work, all LECs entering Eq.~(\ref{lagrange}) should be taken at their
physical values. Further, the SU(2)
matrix $U = u^2$ collects the pion fields  and various covariant derivatives
are defined according to 
\beqa
D_\mu U  &=&  \partial_\mu U -i r_\mu U + i U l_\mu\,, \nn
u_\mu & = & i \left[u^\dagger (\partial_\mu-i r_\mu )u - u (\partial_\mu -i
  l_\mu )u^\dagger   \right] \,,\nn
D_\mu N & = & \left[\partial_\mu + \Gamma_\mu - i v_\mu^{(s)} \right] N \quad 
\mbox{with} \quad
\Gamma_\mu  =  \frac{1}{2}\left[ u^\dagger (\partial_\mu -i r_\mu) u + u
    (\partial_\mu -i l_\mu)u^\dagger\right]\,.
\eeqa
To describe the coupling to an external electromagnetic field, the left- and
right-handed  currents $r_\mu$ and $l_\mu$ and the isoscalar current
$v_\mu^{(s)}$ have to be chosen as
\beq
  r_\mu=l_\mu=\frac{e}{2}\mathcal{A}_\mu \tau_3\,,  \qquad
  v_\mu^{(s)}=\frac{e}{2}\mathcal{A}_\mu \,, \nonumber 
\eeq
where $e$ denotes the elementary charge. 
Expanding the various terms in the effective Lagrangian in powers of the pion
field and using the canonical formalism along the lines of
Ref.~\cite{Gerstein:1971fm}, we  
end up with the following interaction terms in the Hamilton density 
\beqa
\label{vertices}
\mathcal{H}_{21}^{(1)} &=&  \frac{g_A}{2F_\pi} N^\dagger \left(\vec{\sigma}
      \vec{\tau}\cdot\cdot\vec{\nabla}\vec{\pi} \right) N \,, \nn
\mathcal{H}_{22}^{(2)} &=&  \frac{1}{4 F_\pi^2}  N^\dagger
  \big[\vec{\pi}\times\dot{\vec{\pi}} \big]\cdot \vec{\tau} N\,, \nn
\mathcal{H}_{42}^{(4)} &=&  \frac{1}{32 F_\pi^4} \left(N^\dagger
    \left[\vec{\tau}\times\vec{\pi} \right] N \right) \cdot \left(N^\dagger
    \left[\vec{\tau}\times\vec{\pi} \right] N \right)
\,, 
\eeqa
and the electromagnetic current density is of the form
\beqa
{J^0_{20}}^{(-1)} &=&\frac{e}{2} N^\dagger \left(\one
      +\tau_3 \right)N \,, \nn
{J_{02}^0}^{(-1)} &=& e  \big[\vec{\pi}
    \times  \dot{\vec{\pi}} \big]_3 \,, \nn
{\vec J_{02}\,}^{(-1)} &=& - e \big[\vec{\pi}
    \times \vec{\nabla} \vec{\pi} \big]_3 \,, \nn
{\vec J_{21}\,}^{(0)} &=& e\,\frac{
    g_A }{2F_\pi}  N^\dagger         \vec{\sigma} \left[ \vec{\tau} \times \vec{\pi}\right]_3
    N\,.
\eeqa
In the above expressions we adopt the notation of
Ref.~\cite{Epelbaum:2007us}. In particular, the subscripts $a$ and $b$ in
$\mathcal{H}_{ab}^{(\kappa)}$ and ${J^\mu_{ab}}^{(\kappa)}$ refer to the
number of the nucleon and pion fields, respectively, while the
superscript $\kappa$ gives the dimension of the operator as defined in
Eq.~(\ref{pow_fin}). Further, the symbol $\cdot \cdot$ in Eq.~(\ref{vertices})
denotes a scalar product in the spin and isospin spaces. 

\begin{figure}[tb]
\vskip 1 true cm
  \includegraphics[width=11cm,keepaspectratio,angle=0,clip]{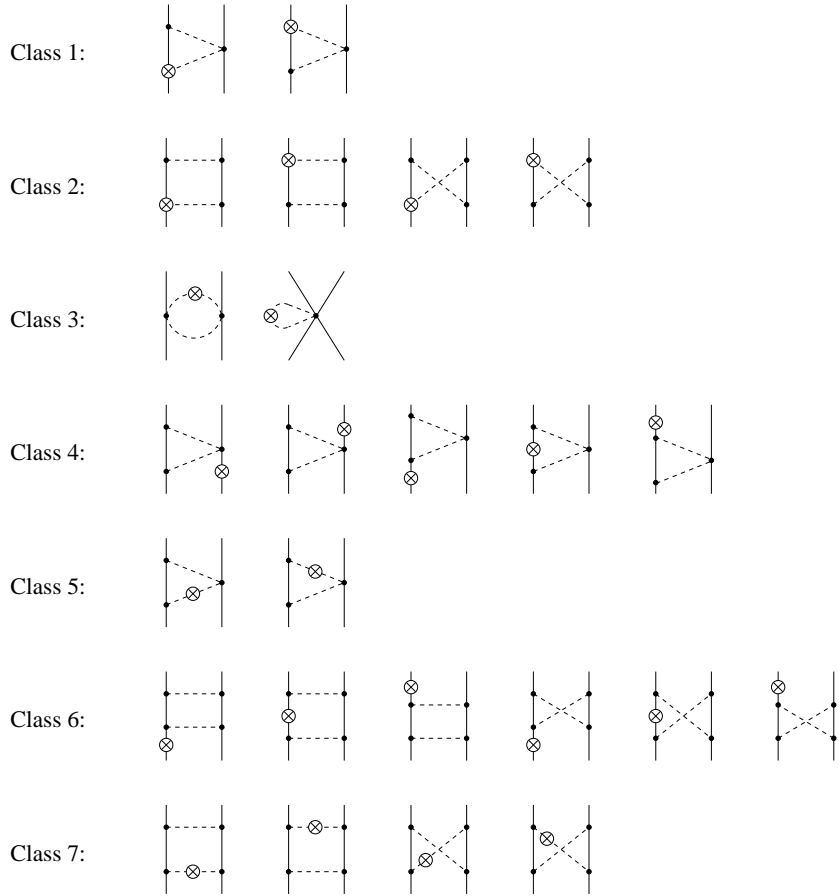}
  \caption{Diagrams showing contributions to the leading two-pion exchange
    currents. Solid and dashed lines refer to nucleons and pions,
    respectively. Solid dots are the lowest-order vertices from the effective
    Lagrangian while the crosses represent insertions of the electromagnetic
    vertices as explained in the text. Diagrams resulting from interchanging
    the nucleon lines are not shown. }
  \label{fig:currents}
\end{figure}

The formal operator structure of the leading two-pion exchange two-nucleon
current at order $\mathcal{O} (e Q )$ is given in appendix \ref{app1}. For the
sake of convenience, we 
distinguish between seven classes of contributions according to the power of
the LEC $g_A$ (i.e.~proportional to $g_A^0$, $g_A^2$ and $g_A^4$) and the type of the
hadronic current $J^\mu_{20}$, $J^\mu_{21}$ or $J^\mu_{02}$ as shown in
Fig.~\ref{fig:currents}. Notice that there are no contributions proportional
to $g_A^0$ and involving $J^\mu_{20}$ and $J^\mu_{21}$. We also emphasize that
the second diagram in the class 3 does not generate any %long-range
contribution. It results from the term in the Hamilton density which is absent
in the Lagrangian and arises through the application of the canonical
formalism.   Finally, it should be understood that the meaning of diagrams
in the method of unitary transformation is different from the one arising in
the context of covariant and/or time-ordered perturbation theory. 
The diagrams shown in Fig.~\ref{fig:currents} serve merely to visualize the
topology corresponding to a given sequence of operators $H$ and $J$ appearing
in the formal expressions given in appendix \ref{app1}.

It is a fairly straightforward albeit tedious exercise
to evaluate the contributions to the nuclear current corresponding to the
operators given  in appendix \ref{app1}. Below, we give explicit results for
the current and charge densities, $J^\mu = (\rho, \, \vec J )$,  resulting from the individual
classes using the notation 
\beq
\label{notation}
\langle \vec p_1 {}'\, \vec p_2 {}' | J^\mu | \vec p_1 \, \vec p_2 \rangle 
= \delta (\vec p_1 {}' + \vec p_2 {}' - \vec p_1 - \vec p_2 - \vec k ) \left[
  \sum_{X=\rm c1}^{\rm c7} J_{\rm X}^\mu + (1 \leftrightarrow 2)\right] \,, 
\eeq
%\beq
%\label{notation}
%\vec{J} = \sum_{X=\rm c1}^{\rm c7} \vec{J}_{\rm X} + (1 \leftrightarrow 2) \,, 
%\quad \quad
%\rho  = \sum_{X=\rm c1}^{\rm c7} \rho_{\rm X} + (1 \leftrightarrow 2)\,,
%\eeq
where $\vec p_i$ ($\vec p_i {} '$) denotes the incoming (outgoing) momentum of
nucleon $i$ and $\vec k$ is the photon momentum.   
Further,  $(1 \leftrightarrow 2)$ refers to the contribution resulting from the
interchange of the nucleon labels. 
We find the following results for the current density
from the individual classes of the diagrams shown in Fig.~\ref{fig:currents}:
\beqa
  \vec{J}_{\rm c1} \left(\vec{q}_{1},\vec{q}_{2} \right) & = & e\,\frac{g_A^2
    \, i \,}{16 F_\pi^4} \left[\vec{q}_1 
   \left[\vec{\tau}_1\times \vec{\tau}_2\right]^3 + 2\left[\vec{q}_1
     \times\vec{\sigma}_2\right]\tau_1^3\right]\int \frac{d^3l}{(2\pi)^3} 
 \frac{1}{\omega_+ \omega_-(\omega_+ + \omega_-)}\nn
& \to &  - e\,\frac{g_{A}^2 \, i \,}{64  \pi^2 F_\pi^4}\biggl[\vec{q}_1
     \left[\vec{\tau}_1\times \vec{\tau}_2\right]^3+2
     \left[\vec{q}_1 \times \vec{\sigma}_2\right]
     \tau_1^3\biggr]L(q_1)\,,\nn
  \vec{J}_{\rm c2} \left(\vec{q}_{1},\vec{q}_{2} \right) & = & 
e\,\frac{g_A^4 \, i \,}{16 F_\pi^4}\int \frac{d^3l}{(2\pi)^3}
\frac{\omega_+^2+\omega_+\omega_-+\omega_-^2}{\omega_+^3\omega_-^3\left(\omega_++\omega_-\right)}\biggl[
  4 \tau_2^3 \ \vec{l} \
  (\vec{l}\cdot\left[\vec{q}_1\times\vec{\sigma}_1\right])
 -  (l^2-q_1^2) \left(\vec{q}_1  \left[\vec{\tau}_1\times \vec{\tau
     }_2\right]^3 + 2 \left[\vec{q}_1\times \vec{\sigma}_2\right]
   \tau_1^3\right)\biggr] \nn
  & \to & e\, \frac{g_A^4 \, i \,}{64  \pi^2 F_\pi^4 \left(q_1^2+4M_{\pi}^2 \right)} \biggl[ \vec{q}_1
  \left[\vec{\tau}_1\times \vec{\tau 
     }_2\right]^3 \left(8 M_\pi^2+3q_1^2\right)+ 2 \left[\vec{q}_1\times
     \vec{\sigma}_2\right] \tau_1^3 \left(8 M_\pi^2+3q_1^2\right)  \nn
&& {} 
 - 4
 \left(q_1^2+4M_{\pi}^2\right) \left[\vec{q}_1\times\vec{\sigma}_1 \right] \tau_2^3\biggr] L(q_1)\,, \nn
  \vec{J}_{\rm c3} \left(\vec{q}_{1},\vec{q}_{2} \right) & = & 
 e\,\frac{i \,}{128 F_\pi^4}\, \left[\vec{\tau}_1\times\vec{\tau}_2\right]\,
\int \frac{d^3l}{(2\pi)^3}
\frac{1}{\omega_1\omega_2\omega_3}\left(\frac{(\omega_1-\omega_2)
    (\omega_1-\omega_3)}{(\omega_1+\omega_2) 
   (\omega_1+\omega_3)}  +  \frac{(\omega_1+\omega_2)
   (\omega_1-\omega_3)}{(\omega_1+\omega_3)
   (\omega_2+\omega_3)}+\frac{(\omega_1-\omega_2)
   (\omega_1+\omega_3)}{(\omega_1+\omega_2) (\omega_2+\omega_3)}\right)\nn
&& {} \times
\left(\vec{k}_2 + \vec{k}_3
 \right) \,, \nn
\vec{J}_{\rm c4} \left(\vec{q}_{1},\vec{q}_{2} \right) & = & \vec{J}_{\rm c6}
\left(\vec{q}_{1},\vec{q}_{2} \right)  = 0\,, \nn
\vec{J}_{\rm c5} \left(\vec{q}_{1},\vec{q}_{2} \right) & = &  -e\,
\frac{g_A^2 \,i}{8 F_\pi^4} \int \frac{d^3l}{(2\pi)^3} \frac{1}{\omega_1\omega_2\omega_3}
\frac{\omega_1+\omega_2+\omega_3}{(\omega_1+\omega_2)(\omega_1+\omega_3)(\omega_2+\omega_3)} 
\left(\vec{k}_{2} + \vec{k}_{3}\right)\biggl[\left[\vec{\tau}_1\times\vec{\tau}_2\right]^3
  \vec{k}_{1}\cdot\vec{k}_{3} \nn 
 &&{}-  2\tau_1^3 
    \vec{\sigma}_2\cdot\left[
      \vec{k}_{1}\times\vec{k}_{3}\right]\biggr]
\,, \nn
\vec{J}_{\rm c7} \left(\vec{q}_{1},\vec{q}_{2} \right) & = & {}e\,
\frac{g_A^4\, i }{16 F_\pi^4 }   \int\frac{d^3l}{(2\pi)^3}  \frac{1}{\omega_1\omega_2\omega_3}\left(
    \frac{1}{\omega_1\omega_2\omega_3^2}-\frac{\omega_2}{\omega_1^2(\omega_1^2-\omega_2^2)
      (\omega_1+\omega_3)}+\frac{\omega_1}{\omega_2^2(\omega_1^2-\omega_2^2)
      (\omega_2+\omega_3)}\right)\nn  
&&{} \times  
\left(\vec{k}_{2}+\vec{k}_{3}\right)
\biggl[\left[\vec{\tau}_1\times\vec{\tau}_2\right]^3
    \vec{k}_{1}\cdot\vec{k}_{2} \,
    \vec{k}_{1}\cdot\vec{k}_{3}
 +  4\tau_2^3 
    \vec{\sigma}_1\cdot\left[
      \vec{k}_{1}\times\vec{k}_{2}\right] \,
    \vec{k}_{1}\cdot\vec{k}_{3} 
% \nn
% &&{}
% -  2\tau_1^3 
%     \vec{\sigma}_2\cdot\left[
%       \vec{k}_{1}\times\vec{k}_{3}\right] \,
%     \vec{k}_{1}\cdot\vec{k}_{2} 
\biggr]  \,, 
\eeqa
while the contributions to the exchange charge density read
\beqa
  \rho_{\rm c1} \left(\vec{q}_{1},\vec{q}_{2} \right) & = & 
\rho_{\rm c2} \left(\vec{q}_{1},\vec{q}_{2} \right) = 
\rho_{\rm c3} \left(\vec{q}_{1},\vec{q}_{2} \right) = 0\,,\nn
\rho_{\rm c4} \left(\vec{q}_{1},\vec{q}_{2} \right) &=& e\,\frac{g_A^2 \,}{16
  F_\pi^4} \tau_1^3 \int \frac{d^3l}{(2\pi)^3} 
\frac{1}{\omega_+^2 \omega_-^2}\left[l^2-q_1^2 \right] \nn
&\to &  -e\,\frac{ {g}_{ A}^2 \, }{32  \pi F_\pi^4} \tau_1^3 \left[M_\pi+ \left(2
      M_\pi^2+q_1^2\right)A(q_1)\right] \,, \nn
\rho_{\rm c5} \left(\vec{q}_{1},\vec{q}_{2} \right) &=&-e\,\frac{g_A^2 }{16
  F_\pi^4}\tau_2^3\int\frac{d^3l}{(2\pi)^3}\frac{1}{\omega_+^2\omega_-^2}(l^2-q_1^2)
\nn
&\to &  e\,\frac{g_A^2 \, }{32
\pi  F_\pi^4}\tau_2^3\left[M_\pi + \left(2M_\pi^2 + q_1^2 \right) A(q_1)\right]\,, \nn
\rho_{\rm c6} \left(\vec{q}_{1},\vec{q}_{2} \right) &=& -e\,\frac{g_A^4 \,}{8 F_\pi^4}\int
\frac{d^3l}{(2\pi)^3}\frac{1}{\omega_+^2\omega_-^4} \biggl[4 \tau_2^3
 \vec{l}\cdot\left[\vec{q}_1\times\vec{\sigma}_1\right]\vec{l}\cdot\left[\vec{q}_1\times
   \vec{\sigma}_2\right] + (l^2-q_1^2)^2\tau_1^3 \biggr]\nn
& \to &  e\,\frac{g_A^2\, }{32  \pi F_\pi^4}\biggl[\biggl(\left(4M_\pi^2+2q_1^2
\right)\tau_1^3
 + \tau_2^3\left[
   \vec{q}_1\cdot\vec{\sigma}_1\vec{q}_1\cdot\vec{\sigma}_2-q_1^2
   \vec{\sigma}_1\cdot\vec{\sigma}_2\right] \biggr) 
A(q_1) + M_\pi\frac{\left(11M_\pi^2+ 3q_1^2
  \right)}{4M_\pi^2+q_1^2} \tau_1^3\biggr]\,,\nn
\rho_{\rm c7} \left(\vec{q}_{1},\vec{q}_{2} \right) &= & {}e\,\frac{g_A^4
 }{8 F_\pi^4} \int \frac{d^3l}{(2\pi)^3}\frac{1}{\omega_1^2 \omega_2^2 \omega_3^2}
\biggl[ 
\left(\tau_1^3 + \tau_2^3\right)
 \biggl(
\vec{k}_{1}\cdot\vec{k}_{2} \vec{k}_{1}\cdot\vec{k}_{3}
+ \vec{k}_1\cdot\left[\vec{k}_2\times\vec{\sigma}_1\right] \vec{k}_1\cdot\left[\vec{k}_3\times\vec{\sigma}_2\right]
\biggr) \nn 
&&\qquad 
{}+\left[\vec{\tau}_1\times\vec{\tau}_2\right]^3
\vec{k}_{1}\cdot\vec{k}_{2}\vec{k}_{1}\cdot\left[\vec{k}_{3}\times\vec{\sigma}_2\right]
 \biggr] 
 \,,
\eeqa
where $\vec q_i \equiv \vec p_i \, ' - \vec p_i$ is the momentum transfer of
nucleon $i$. Clearly, $\vec \sigma_i$ and $\vec \tau_i$ refer to
the spin and isospin Pauli matrices of the nucleon $i$. Here and in what
follows, we label the Cartesian components of various
vectors in isospin space by the superscripts rather than subscripts in order
to avoid a possible confusion with the nucleon labels. 
Further, $\omega_\pm =\sqrt{ (\vec{l} + \vec{q}_1
 )^2 + 4M_\pi^2}$ and $\omega_i = \sqrt{\vec k_i^2 + M_\pi^2}$ with 
$\vec k_1 = \vec l$, $\vec k_2 
= \vec l - \vec q_1$ and $\vec k_3 = \vec l + \vec q_2$. We also use here the
short-hand notation with $q_i \equiv | \vec q_i |$. Finally, it should be understood
that the above expressions refer to matrix elements with respect to momenta
and operators in  spin and isospin spaces, see Eq.~(\ref{notation}). 
The expressions for $\vec{J}_{\rm c1,c2} 
\left(\vec{q}_{1},\vec{q}_{2} \right)$ and $\rho_{\rm c4,c5,c6}
\left(\vec{q}_{1},\vec{q}_{2} \right)$ only involve two-point functions so
that the corresponding integrals can be evaluated in terms of the loop
functions $L(q)$ and $A(q)$ defined as 
\begin{eqnarray}
\label{LundA}
  L\left( q \right) & = &
  \frac{1}{2}\frac{s}{q}\ln\left(\frac{s+q}{s-q}\right), \quad \textrm{with}
  \ s = \sqrt{q^2+4M_\pi^2} \,, \nn
A\left( q  \right) & = & \frac{1}{2q}\arctan \left( \frac{q}{2M_\pi}\right) \,.
\end{eqnarray}
Notice that the loop integrals are ultraviolet divergent and
thus need to be renormalized by an appropriate redefinition of the LECs
accompanying the NN
contact interactions. Here we only show the non-polynomial parts of the resulting
expressions which give rise
to long-range contributions and are uniquely defined once the regulator
(i.e. cutoff) is removed. Short-range contributions involving NN contact
interactions will be considered elsewhere. The symbol $\to$ in the above
equations signifies that the original ill-defined expression is evaluated using
dimensional regularization\footnote{Using e.g.~a cutoff regularization will lead to
  the same result after taking the limit $\Lambda \to \infty$.} and only
non-polynomial in momenta and $M_\pi^2$ contributions are shown explicitly. 
The exchange current/charge density
contributions 
$\vec{J}_{\rm c3,c5,c7} \left(\vec{q}_{1},\vec{q}_{2} \right)$ and 
$\rho_{\rm c7} \left(\vec{q}_{1},\vec{q}_{2} \right)$
result from
loop diagrams where a photon couples to pions in flight. The corresponding
loop integrals depend explicitly on two external momenta $\vec q_1$ and $\vec
q_2$ and can be written in terms of the three-point functions. The complete
expressions in momentum space are given in appendix \ref{app2} in terms of a
number of scalar integrals, which are easily calculable numerically as
detailed in appendix \ref{app3}. We also emphasize that we have explicitly
verified in appendix \ref{app4} that the derived exchange currents fulfill the
continuity equation. Notice further that all loop
integrals can be carried out analytically in configuration space as discussed
in the next section. 
The results for the current density presented here agree with the ones 
of~\cite{Pastore:2009is}.\footnote{For contributions involving the three-point
function we could only check the 
intermediate results for the integrands involving different pion energies.} 
%There are, however, differences in $\vec{J}_{\rm
%  c1}$ and $\vec{J}_{\rm c5}$, corresponding to formulae (5.1) and (5.2) in
%the cited paper. In both formulae, the isospin operator $\tau_{2,z}$ has to be
%multiplied by a factor of 2 so that the results agree with ours.   

%%%%%%%%%%%%%%%%%%%%%%%%%%%%%%%%%%%%%%%%%%%%%%%%%%%%%%%%%%%%%%%%%%%%%%%%%%%%%%%%%
\section{Two-pion exchange current in configuration space}
\def\theequation{\arabic{section}.\arabic{equation}}
\setcounter{equation}{0}
\label{sec2}

The obtained expressions in momentum space depend only on the momentum transfers of the individual
nucleons leading to a local form of the current operator in configuration
space which is defined according to 
\begin{eqnarray}
  J^\mu \left(\vec{r}_{10},\vec{r}_{20} \right) & \equiv & 
\int \frac{d^3q_1}{(2\pi)^3}\frac{d^3q_2}{(2\pi)^3}e^{i
 \vec{q}_1 \cdot \vec{r}_{10}}\; e^{i
  \vec{q}_2 \cdot\vec{r}_{20}}J^\mu (\vec{q}_1,\vec{q}_2),
\end{eqnarray}
where $\vec{r}_{10}=\vec{r}_1-\vec{r}_0$, $\vec{r}_{20}=\vec{r}_2-\vec{r}_0$
and $\vec{r}_1$, $\vec{r}_2$ and $\vec{r}_0$ denote the positions
of the nucleons 1 and 2 and the photon coupling, respectively.  
Using the formulae collected in appendix
\ref{app5} and expressions for the current operator in momentum space in terms of
three-dimensional loop integrals given in the previous section, we obtain the
following surprisingly compact expressions for the current density

\beqa
\vec{J}_{\rm c1}\left(\vec{r}_{10},\vec{r}_{20} \right) & = & {}e\,\frac{g_A^2
  \, M_\pi^7 \,}{128  \pi^3 F_\pi^4}\left[\vec{\nabla}_{10} 
   \left[\vec{\tau}_1\times \vec{\tau}_2\right]^3 + 2\left[\vec{\nabla}_{10}
     \times\vec{\sigma}_2\right]\tau_1^3\right]\, \delta (\vec x_{20}) \,
 \frac{K_1(2x_{10})}{x_{10}^2}
\,,\nn
\vec{J}_{\rm c2}\left(\vec{r}_{10},\vec{r}_{20} \right) & = &
{}-e\,\frac{g_A^4 \, M_\pi^7 \,}{256 \pi^3 F_\pi^4 } 
 \left(3\nabla_{10}^2-8 \right) \, \biggl[
  \vec{\nabla}_{10}\left[\vec{\tau}_1\times \vec{\tau}_2\right]^3  + 2 \left[
    \vec{\nabla}_{10} \times \vec{\sigma}_2 \right]\tau_1^3\biggr]\,
  \delta (\vec x_{20}) \, \frac{K_0(2
  x_{10})}{x_{10}} \nn
&&{}+ e \,\frac{g_A^4 \, M_\pi^7 \,}{32 \pi^3  F_\pi^4 } \left[
    \vec{\nabla}_{10} \times \vec{\sigma}_1 \right]\tau_2^3\,
  \delta (\vec x_{20}) \, \frac{K_1(2
  x_{10})}{x_{10}^2}
\,, \nn
\vec{J}_{\rm c3}\left(\vec{r}_{10},\vec{r}_{20} \right) & = & {}-e\,\frac{M_\pi^7}{512
     \pi^4 F_\pi^4
   }\left[\vec{\tau}_1\times\vec{\tau}_2\right]^3(\vec{\nabla}_{10}
  -\vec{\nabla}_{20})\frac{K_2(x_{10}   
    + x_{20} + x_{12})}{(x_{10} \, x_{20} \, x_{12})(x_{10} + x_{20} +
    x_{12})} \,, \nn
\vec{J}_{\rm c5} \left(\vec{r}_{10},\vec{r}_{20} \right) & = &{} -e\,
\frac{g_A^2 \, M_\pi^7}{256  \pi^4 F_\pi^4} 
\left( \vec{\nabla}_{10} - \vec{\nabla}_{20}\right)\biggl[\left[\vec{\tau}_1\times\vec{\tau}_2\right]^3
\vec{\nabla}_{12}\cdot\vec{\nabla}_{20} 
 -  2\tau_1^3 
    \vec{\sigma}_2\cdot\left[
      \vec{\nabla}_{12}\times\vec{\nabla}_{20}\right]\biggr]\nn
 &&{}\times \frac{K_1 (x_{10}
      + x_{20} + x_{12})}{(x_{10} \, x_{20} \, x_{12})} 
\,, \nn
\vec{J}_{\rm c7}\left(\vec{r}_{10},\vec{r}_{20} \right)  & = &{}e\,
\frac{g_A^4 \, M_\pi^7}{512  \pi^4 F_\pi^4}  
\left( \vec{\nabla}_{10}-\vec{\nabla}_{20}\right)\biggl[\left[\vec{\tau}_1\times\vec{\tau}_2\right]^3
    \vec{\nabla}_{12}\cdot\vec{\nabla}_{10} \,
    \vec{\nabla}_{12}\cdot\vec{\nabla}_{20}
 +  4\tau_2^3 
    \vec{\sigma}_1\cdot\left[
      \vec{\nabla}_{12}\times\vec{\nabla}_{10}\right] \,
    \vec{\nabla}_{12}\cdot\vec{\nabla}_{20} 
\biggr] 
    \nn
&&{} \times \frac{x_{10} + x_{20} + x_{12}}{x_{10} \, x_{20} \, x_{12}} K_0(x_{10} +
x_{20} + x_{12})
\,, 
\eeqa
and the charge density 
\beqa
\rho_{\rm c1}\left(\vec{r}_{10},\vec{r}_{20} \right)  & = &{} \rho_{\rm
  c2}\left(\vec{r}_{10},\vec{r}_{20} \right) = \rho_{\rm
  c3}\left(\vec{r}_{10},\vec{r}_{20} \right)  = 0\,,    \nn
\rho_{\rm c4}\left(\vec{r}_{10},\vec{r}_{20} \right)  & = &{} e\,\frac{g_A^2
  \,M_\pi^7}{256  \pi^2 F_\pi^4} \, 
\tau_1^3 \, \delta(\vec x_{20}) \, \left(\nabla_{10}^2 - 2 \right)
\frac{e^{-2x_{10}}}{x_{10}^2}
 \,,\nn
\rho_{\rm c5}\left(\vec{r}_{10},\vec{r}_{20} \right)  & = &{}- e\,\frac{g_A^2
  \,M_\pi^7}{256 \pi^2  F_\pi^4} \, 
\tau_2^3 \, \delta (\vec x_{20}) \, \left(\nabla_{10}^2 - 2 \right)
 \frac{e^{-2x_{10}}}{x_{10}^2}
 \,,\nn
\rho_{\rm c6}\left(\vec{r}_{10},\vec{r}_{20} \right)  & = & {}- 
{}e\,\frac{g_A^4 \,M_\pi^7}{256 \pi^2 F_\pi^4 } \, \delta (\vec x_{20}) \, \biggl[ \tau_1^3 \left(2
  \nabla_{10}^2 -4\right) + \tau_2^3 \, \vec{\sigma}_1 \cdot \vec{\nabla}_{10}
\, \vec{\sigma}_2 \cdot \vec{\nabla}_{10} \, - \tau_2^3 \vec{\sigma}_1 \cdot
\vec{\sigma}_2\biggr]\frac{e^{-2x_{10}}}{x_{10}^2}\nn
 &&{} - e\,\frac{g_A^4 \,M_\pi^7}{128 \pi^2 F_\pi^4} \, \delta (\vec x_{20}) \,\tau_1^3 \,
 \left(3 \nabla_{10}^2 -11 \right) \, \frac{e^{-2x_{10}}}{x_{10}}  
 \,,\nn
\rho_{\rm c7} \left(\vec{r}_{10},\vec{r}_{20} \right) & = & {}-e\,\frac{g_A^4
  \,M_\pi^7}{512 \pi^3 F_\pi^4 }
\biggl[ 
(
\tau_1^3
+\tau_2^3)
 \biggl(\vec{\nabla}_{12}\cdot\vec{\nabla}_{10}
   \vec{\nabla}_{12}\cdot\vec{\nabla}_{20} +
   \vec{\nabla}_{12}\cdot\left[\vec{\nabla}_{10}\times\vec{\sigma}_1\right]
\vec{\nabla}_{12}\cdot\left[\vec{\nabla}_{20}\times\vec{\sigma}_2\right] \biggr)\nn
&&{}+ \left[\vec{\tau}_1\times\vec{\tau}_2\right]^3
\vec{\nabla}_{12}\cdot\vec{\nabla}_{10} \, 
  \vec{\nabla}_{12}\cdot\left[\vec{\nabla}_{20}\times\vec{\sigma}_2\right]
  \biggr]\, 
\frac{e^{-x_{10}}}{x_{10}}\;\frac{e^{-x_{20}}}{x_{20}}\;\frac{e^{-x_{12}}}{x_{12}}
 \,.
\eeqa
In the above expressions, $K_n(x)$ denote the modified Bessel functions of the
second kind and we have introduced dimensionless variables $\vec
x_{10} = M_\pi \vec r_{10}$, $\vec x_{20} = M_\pi \vec r_{20}$ and  
$\vec x_{12} = M_\pi \vec r_{12} = M_\pi  (\vec r_1 - \vec r_2 \, )$. 
Further,
$x_{ij} \equiv | \vec x_{ij}  |$ and all derivatives with respect to $\vec
x_{10}$, $\vec x_{20}$
and $\vec x_{12}$ are to be evaluated as if these variables were independent of
each other. We also emphasize that the above expressions are valid for $x_{10} +
x_{20} > 0$. Finally, it should be understood that the behavior of the current
and charge densities at short distances will be affected if one uses a
regularization with a finite value of the cutoff.

%%%%%%%%%%%%%%%%%%%%%%%%%%%%%%%%%%%%%%%%%%%%%%%%%%%%%%%%%%%%%%%%%%%%%%%%%%%%%%%%%
\section{Summary and outlook}
\def\theequation{\arabic{section}.\arabic{equation}}
\label{sec:summary}

In this paper, we applied the method of unitary transformation to derive
the leading two-pion exchange two-nucleon charge and current densities based
on chiral effective field theory. The resulting nuclear current is given both
in momentum and configuration space. The results in momentum space involve
the standard loop functions $L(q)$ and $A(q)$ and, for
certain classes of diagrams, also the three-point functions. In
the latter case we expressed all tensor integrals in terms of a set of scalar
integrals in different dimensions similar to the method of 
Ref.~\cite{Davydychev:1991va}, see appendix~\ref{app2}. For the remaining
scalar integrals we 
derive in appendix \ref{app3} a simple representation using Feynman
parameters which is well suited for numerical calculations. In configuration
space, we were able to evaluate all loop integrals analytically leading to very
compact expressions in terms of the modified Bessel functions of the second kind. We have
also explicitly verified that the derived exchange currents fulfill the
continuity equation, see appendix \ref{app4}. 

In addition to the two-pion exchange contributions, there are also 
one-pion exchange and short-range terms at order $\mathcal{O}
(eQ)$, see Refs.~\cite{Pastore:2008ui,Pastore:2009is} for a recent work based
on time-ordered perturbation 
theory. As will be demonstrated in a subsequent publication
\cite{Koelling:future}, renormalization 
of the one-pion exchange contributions at the one-loop level strongly
restricts the ambiguity in the definition of the current
and provides a highly non-trivial consistency check of the calculation. In
particular, one needs to ensure that \emph{all} appearing ultraviolet
divergences are absorbed into redefinition of the LECs $d_i$ and $l_i$ from
$\mathcal{L}_{\pi N}^{(3)}$ and $\mathcal{L}_{\pi}^{(4)}$, respectively, with
already known $\beta$-functions, see
e.g.~\cite{Gasser:1984ux,Ecker:1995rk,Fettes:1998ud,Gasser:2002am}. This work
is in progress \cite{Koelling:future}. 

Finally, in the future, one also needs to test the convergence of the chiral
expansion for the one- and two-pion exchange currents by calculating the
corrections at order $\mathcal{O} (e Q^2)$. Given the large numerical values
of the LECs $c_{3,4}$ from $\mathcal{L}_{\pi N}^{(2)}$, one might expect
sizeable corrections which, indeed, is well known to be the case for the
two-pion exchange potential \cite{Kaiser:1997mw}. In this context, it might be advantageous
to include the $\Delta$(1232) isobar as an explicit degree of freedom in
effective field theory utilizing the small scale expansion
\cite{Hemmert:1997ye}, see
\cite{Ordonez:1995rz,Kaiser:1998wa,Krebs:2007rh,Epelbaum:2007sq,Epelbaum:2008td,Pandharipande:2005sx} for recent
work along these lines in the purely strong few-nucleon sector.

%%%%%%%%%%%%%%%%%%%%%%%%%%%%%%%%%%%%%%%%%%%%%%%%%%%%%%%%%%%%%%%%%%%%%%%%%%%%%%%%%
\section*{Acknowledgments}

We would like to thank  Daniel Phillips for useful comments on the
manuscript and Walter Gl\"ockle, Jacek Golak, Dagmara Rozpedzik and
Henryk Wita{\l}a for many stimulating discussions on this topic.  S.K.~would
also like to thank Jacek Golak and Henryk Wita{\l}a for their hospitality
during his stay in Krakow where a part of this work was done. 
This work was
supported by funds provided by the Helmholtz Association to the
young investigator group ``Few-Nucleon Systems in Chiral Effective
Field Theory'' (grant VH-NG-222) and to the virtual institute ``Spin
and strong QCD'' (VH-VI-231), by the DFG (SFB/TR 16 ``Subnuclear
Structure of Matter'') and by the EU HadronPhysics2 project ``Study
of strongly interacting matter''.

%%%%%%%%%%%%%%%%%%%%%%%%%%%%%%%%%%%%%%%%%%%%%%%%%%%%%%%%%%%%%%%%%%%%%%%%%%%%%%%%%
\appendix

\def\theequation{\Alph{section}.\arabic{equation}}
\setcounter{equation}{0}
\section{Formal structure of the leading two-pion exchange current operator}
\label{app1}
%%%%%%%%%%%%%%%%%%%%%%%%%%%%%%%%%%%%%%%%%%%%%%%%%%%%%%%%%%%%%%%%%%%%%%%%%%%%%%%%%

In this appendix we give the formal structure of the two-pion
exchange current operator that results after applying the unitary
transformation. All vertices entering the expressions below should be
understood as second-quantized normally-ordered operators. Here and in what
follows, we adopt the notation of Ref.~\cite{Epelbaum:2007us} with a few minor
modifications. 

\begin{itemize}
\item\textbf{Class 1} contributions involving $J_{21}^{(0)}$ and proportional to $g_A^2$
\beq
\label{eq:EffectiveCurrentNLO1}
    J_{\rm c1} = 
         \eta  \left[  H_{22}^{(2)} \frac{\lambda^2}{E_\pi}
         J_{21}^{(0)} \frac{\lambda^1}{E_\pi} H_{21}^{(1)}  
           +  H_{22}^{(2)} \frac{\lambda^2}{E_\pi} H_{21}^{(1)}
           \frac{\lambda^1}{E_\pi} J_{21}^{(0)}  
        +  \ J_{21}^{(0)} \frac{\lambda^1}{E_\pi} H_{22}^{(2)}
        \frac{\lambda^1}{E_\pi} H_{21}^{(1)} \right] \eta + \text{h.c.}\,.
  \eeq    
\item\textbf{Class 2} contributions involving $J_{21}^{(0)}$ and proportional to
  $g_A^4$
\begin{eqnarray}
\label{eq:EffectiveCurrentNLO2}
      J_{\rm c2} &=& \eta \left[ J_{21}^{(0)} \frac{\lambda^1}{E_\pi^2}
        H_{21}^{(1)} \eta  H_{21}^{(1)} \frac{\lambda^1}{E_\pi} H_{21}^{(1)}   
        + \frac{1}{2}   J_{21}^{(0)} \frac{\lambda^1}{E_\pi} H_{21}^{(1)} \eta
        H_{21}^{(1)} \frac{\lambda^1}{E_\pi^2} H_{21}^{(1)} -   J_{21}^{(0)} \frac{\lambda^1}{E_\pi} H_{21}^{(1)}
      \frac{\lambda^2}{E_\pi} H_{21}^{(1)} \frac{\lambda^1}{E_\pi} H_{21}^{(1)}  \right. \nn
      & &{}
      + \frac{1}{2}   H_{21}^{(1)} \frac{\lambda^1}{E_\pi^2} H_{21}^{(1)} \eta
      J_{21}^{(0)} \frac{\lambda^1}{E_\pi} H_{21}^{(1)}   - \left.  H_{21}^{(1)} \frac{\lambda^1}{E_\pi} J_{21}^{(0)}
        \frac{\lambda^2}{E_\pi} H_{21}^{(1)} \frac{\lambda^1}{E_\pi} H_{21}^{(1)} \right] \eta + \text{h.c.}\,.
\end{eqnarray}
\item\textbf{Class 3} contributions involving $J_{02}^{(-1)}$ and proportional to $g_A^0$
\beq
\label{eq:EffectiveCurrentNLO3}
      J_{\rm c3} = \eta \left[ H_{22}^{(2)} \frac{\lambda^2}{E_\pi} H_{22}^{(2)} \frac{\lambda^2}{E_\pi} J_{02}^{(-1)}   
        + \frac{1}{2}  H_{22}^{(2)} \frac{\lambda^2}{E_\pi}
        J_{02}^{(-1)}\frac{\lambda^2}{E_\pi} H_{22}^{(2)}- J_{02}^{(-1)}
        \frac{\lambda^2}{E_\pi} H_{42}^4 \right]\eta+  \text{h.c.}\,. 
\eeq
\item\textbf{Class 4} contributions involving $J_{20}^{(-1)}$ and proportional to $g_A^2$
 \begin{eqnarray}
\label{eq:EffectiveCurrentNLO4}
      J_{\rm c4} &=&  \eta \left[- H_{22}^{(2)} \frac{\lambda^2}{E_\pi}
        J_{20}^{(-1)} \frac{\lambda^2}{E_\pi} H_{21}^{(1)} \frac{\lambda^1}{E_\pi}
        H_{21}^{(1)}   
        + \frac{1}{2}   H_{22}^{(2)} \frac{\lambda^2}{E_\pi} H_{21}^{(1)}
        \frac{\lambda^1}{E_\pi^2} H_{21}^{(1)} \eta  J_{20}^{(-1)}  \right. 
      -    H_{22}^{(2)} \frac{\lambda^2}{E_\pi} H_{21}^{(1)}
      \frac{\lambda^1}{E_\pi} J_{20}^{(-1)} \frac{\lambda^1}{E_\pi} H_{21}^{(1)}   \nn
      &&{} +   J_{20}^{(-1)} \eta  H_{22}^{(2)} \frac{\lambda^2}{E_\pi^2} H_{21}^{(1)}
      \frac{\lambda^1}{E_\pi} H_{21}^{(1)}  
      +  \frac{1}{2}   J_{20}^{(-1)} \eta  H_{22}^{(2)} \frac{\lambda^2}{E_\pi}
      H_{21}^{(1)} \frac{\lambda^1}{E_\pi^2} H_{21}^{(1)}   
      + \frac{1}{2}   J_{20}^{(-1)} \eta  H_{21}^{(1)} \frac{\lambda^1}{E_\pi^2}
      H_{22}^{(2)} \frac{\lambda^1}{E_\pi} H_{21}^{(1)}  \nn 
      & &{} \left.{} +  \frac{1}{2}   J_{20}^{(-1)} \eta  H_{21}^{(1)}
        \frac{\lambda^1}{E_\pi} H_{22}^{(2)} \frac{\lambda^1}{E_\pi^2}
        H_{21}^{(1)}   
        -   H_{21}^{(1)} \frac{\lambda^1}{E_\pi} H_{22}^{(2)} 
        \frac{\lambda^1}{E_\pi} J_{20}^{(-1)} \frac{\lambda^1}{E_\pi} H_{21}^{(1)} 
      \right] \eta + \text{h.c.}\,.
 \end{eqnarray}
\item\textbf{Class 5} contributions involving $J_{02}^{(-1)}$ and proportional to $g_A^2$
\begin{eqnarray}
\label{eq:EffectiveCurrentNLO5}
      J_{\rm c5} &= 
      &   \eta \left[ - H_{22}^{(2)} \frac{\lambda^2}{E_\pi}
        J_{02}^{(-1)} \frac{\lambda^2}{E_\pi} H_{21}^{(1)} \frac{\lambda^1}{E_\pi}
        H_{21}^{(1)}   
        -   J_{02}^{(-1)} \frac{\lambda^2}{E_\pi} H_{22}^{(2)}
        \frac{\lambda^2}{E_\pi} H_{21}^{(1)} \frac{\lambda^1}{E_\pi} H_{21}^{(1)}  \right. 
       -   H_{22}^{(2)} \frac{\lambda^2}{E_\pi} H_{21}^{(1)}
      \frac{\lambda^{1}}{E_\pi} H_{21}^{(1)} \frac{\lambda^2}{E_\pi}
      J_{02}^{(-1)}   \nn
     && {}  -    H_{21}^{(1)} \frac{\lambda^1}{E_\pi} H_{22}^{(2)}
      \frac{\lambda^{1}}{E_\pi} J_{02}^{(-1)} \frac{\lambda^1}{E_\pi}
      H_{21}^{(1)}  
-   H_{22}^{(2)} \frac{\lambda^2}{E_\pi} H_{21}^{(1)}
      \frac{\lambda^{3}}{E_\pi} H_{21}^{(1)} \frac{\lambda^2}{E_\pi}
      J_{02}^{(-1)} 
 -    H_{21}^{(1)} \frac{\lambda^1}{E_\pi} H_{22}^{(2)}
      \frac{\lambda^{3}}{E_\pi} J_{02}^{(-1)} \frac{\lambda^1}{E_\pi}
      H_{21}^{(1)}  \nn
      & & {} -    J_{02}^{(-1)} \frac{\lambda^2}{E_\pi} H_{21}^{(1)}
      \frac{\lambda^{1}}{E_\pi} H_{22}^{(2)} \frac{\lambda^1}{E_\pi}
      H_{21}^{(1)}   
      -   H_{22}^{(2)} \frac{\lambda^2}{E_\pi} H_{21}^{(1)}
      \frac{\lambda^{1}}{E_\pi} J_{02}^{(-1)} \frac{\lambda^1}{E_\pi}
      H_{21}^{(1)}
 -    J_{02}^{(-1)} \frac{\lambda^2}{E_\pi} H_{21}^{(1)}
      \frac{\lambda^{3}}{E_\pi} H_{22}^{(2)} \frac{\lambda^1}{E_\pi}
      H_{21}^{(1)}   \nn
      & & {}  -   H_{22}^{(2)} \frac{\lambda^2}{E_\pi} H_{21}^{(1)}
      \frac{\lambda^{3}}{E_\pi} J_{02}^{(-1)} \frac{\lambda^1}{E_\pi}
      H_{21}^{(1)}
       + \frac{1}{2} H_{22}^{(2)}\frac{\lambda^2}{E_\pi}J_{02}^{(-1)}\eta
       H_{21}^{(1)} \frac{\lambda^1}{E_\pi^2} H_{21}^{(1)}  
      + \frac{1}{2}  J_{02}^{(-1)} \frac{\lambda^2}{E_\pi} H_{22}^{(2)}\eta
      H_{21}^{(1)}\frac{\lambda^1}{E_\pi^2} H_{21}^{(1)}\nonumber\\ 
      & & {}\left. {}+ J_{02}^{(-1)}\frac{\lambda^2}{E_\pi^2} H_{22}^{(2)}\eta 
        H_{21}^{(1)}\frac{\lambda^1}{E_\pi} H_{21}^{(1)} \right] \eta  +
      \text{h.c.}
\end{eqnarray}
\item\textbf{Class 6} contributions involving $J_{20}^{(-1)}$ and proportional to $g_A^4$
\begin{eqnarray}
\label{eq:EffectiveCurrentNLO6}
      J_{\rm c6} &=&
  \eta \left[ J_{20}^{(-1)} \eta  H_{21}^{(1)}
        \frac{\lambda^1}{E_\pi^3} H_{21}^{(1)} \eta  H_{21}^{(1)}
        \frac{\lambda^1}{E_\pi} H_{21}^{(1)}   
        + \frac{3}{8}   J_{20}^{(-1)} \eta  H_{21}^{(1)} \frac{\lambda^1}{E_\pi^2}
        H_{21}^{(1)} \eta  H_{21}^{(1)} \frac{\lambda^1}{E_\pi^2} H_{21}^{(1)}
      \right. \nn
      &&{}-\frac{3}{4}   J_{20}^{(-1)} \eta  H_{21}^{(1)} \frac{\lambda^1}{E_\pi^2}
      H_{21}^{(1)} \frac{\lambda^2}{E_\pi} H_{21}^{(1)} \frac{\lambda^1}{E_\pi}
      H_{21}^{(1)}   
      - \frac{1}{2}   J_{20}^{(-1)} \eta  H_{21}^{(1)} \frac{\lambda^1}{E_\pi}
      H_{21}^{(1)} \frac{\lambda^2}{E_\pi^2} H_{21}^{(1)}
      \frac{\lambda^1}{E_\pi} H_{21}^{(1)}  \nonumber\\ 
      & &{} - \frac{1}{4}   J_{20}^{(-1)} \eta  H_{21}^{(1)} \frac{\lambda^1}{E_\pi}
      H_{21}^{(1)} \frac{\lambda^2}{E_\pi} H_{21}^{(1)}
      \frac{\lambda^1}{E_\pi^2} H_{21}^{(1)}   
      + \frac{1}{8}   H_{21}^{(1)} \frac{\lambda^1}{E_\pi^2} H_{21}^{(1)} \eta
      J_{20}^{(-1)} \eta  H_{21}^{(1)} \frac{\lambda^1}{E_\pi^2} H_{21}^{(1)}
      \nonumber\\ 
      & &{} -\frac{1}{2}   H_{21}^{(1)} \frac{\lambda^1}{E_\pi^2} H_{21}^{(1)} \eta
      H_{21}^{(1)} \frac{\lambda^1}{E_\pi} J_{20}^{(-1)} \frac{\lambda^1}{E_\pi}
      H_{21}^{(1)}   
      -   H_{21}^{(1)} \frac{\lambda^1}{E_\pi} J_{20}^{(-1)}
      \frac{\lambda^1}{E_\pi^2} H_{21}^{(1)} \eta  H_{21}^{(1)}
      \frac{\lambda^1}{E_\pi} H_{21}^{(1)}  \nonumber\\ 
      & & {}\left. {}+    H_{21}^{(1)} \frac{\lambda^1}{E_\pi} J_{20}^{(-1)}
        \frac{\lambda^1}{E_\pi} H_{21}^{(1)} \frac{\lambda^2}{E_\pi} H_{21}^{(1)}
        \frac{\lambda^1}{E_\pi} H_{21}^{(1)}    
        + \frac{1}{2}  H_{21}^{(1)} \frac{\lambda^1}{E_\pi} H_{21}^{(1)} 
        \frac{\lambda^2}{E_\pi} J_{20}^{(-1)} \frac{\lambda^2}{E_\pi} H_{21}^{(1)} 
        \frac{\lambda^1}{E_\pi} H_{21}^{(1)} \right] \eta+   \text{h.c.}\,.
 \end{eqnarray}
\item\textbf{Class 7} contributions involving $J_{02}^{(-1)}$ and proportional to $g_A^4$
    \begin{eqnarray}
\label{eq:EffectiveCurrentNLO7}
      J_{\rm c7}&=& 
  \eta \left[- J_{02}^{(-1)} \frac{\lambda^2}{E_\pi^2}
        H_{21}^{(1)} \frac{\lambda^1}{E_\pi} H_{21}^{(1)} \eta  H_{21}^{(1)}
        \frac{\lambda^1}{E_\pi} H_{21}^{(1)}   
        -   J_{02}^{(-1)} \frac{\lambda^2}{E_\pi} H_{21}^{(1)}
        \frac{\lambda^1}{E_\pi^2} H_{21}^{(1)} \eta  H_{21}^{(1)}
        \frac{\lambda^1}{E_\pi} H_{21}^{(1)} \right. \nonumber\\ 
      & & {}-  \frac{1}{2}   J_{02}^{(-1)} \frac{\lambda^2}{E_\pi} H_{21}^{(1)}
      \frac{\lambda^1}{E_\pi} H_{21}^{(1)} \eta  H_{21}^{(1)}
      \frac{\lambda^1}{E_\pi^2} H_{21}^{(1)}   
      +   J_{02}^{(-1)} \frac{\lambda^2}{E_\pi} H_{21}^{(1)}
      \frac{\lambda^{1}}{E_\pi} H_{21}^{(1)} \frac{\lambda^2}{E_\pi}
      H_{21}^{(1)} \frac{\lambda^1}{E_\pi} H_{21}^{(1)}  \nonumber\\ 
      & &{} +  H_{21}^{(1)} \frac{\lambda^1}{E_\pi} J_{02}^{(-1)}
      \frac{\lambda^{1}}{E_\pi} H_{21}^{(1)} \frac{\lambda^2}{E_\pi}
      H_{21}^{(1)} \frac{\lambda^1}{E_\pi} H_{21}^{(1)}   
+   J_{02}^{(-1)} \frac{\lambda^2}{E_\pi} H_{21}^{(1)}
      \frac{\lambda^{3}}{E_\pi} H_{21}^{(1)} \frac{\lambda^2}{E_\pi}
      H_{21}^{(1)} \frac{\lambda^1}{E_\pi} H_{21}^{(1)} \nn
&& {}+  H_{21}^{(1)} \frac{\lambda^1}{E_\pi} J_{02}^{(-1)}
      \frac{\lambda^{3}}{E_\pi} H_{21}^{(1)} \frac{\lambda^2}{E_\pi}
      H_{21}^{(1)} \frac{\lambda^1}{E_\pi} H_{21}^{(1)}   
      - \frac{1}{2}   H_{21}^{(1)} \frac{\lambda^1}{E_\pi^2} H_{21}^{(1)} \eta
      J_{02}^{(-1)} \frac{\lambda^2}{E_\pi} H_{21}^{(1)} \frac{\lambda^1}{E_\pi}
      H_{21}^{(1)}  \nonumber\\ 
      & &{} -  \frac{1}{2}   H_{21}^{(1)} \frac{\lambda^1}{E_\pi^2} H_{21}^{(1)}
      \eta  H_{21}^{(1)} \frac{\lambda^1}{E_\pi} J_{02}^{(-1)}
      \frac{\lambda^1}{E_\pi} H_{21}^{(1)}   
      -    H_{21}^{(1)} \frac{\lambda^1}{E_\pi} J_{02}^{(-1)}
      \frac{\lambda^1}{E_\pi^2} H_{21}^{(1)} \eta  H_{21}^{(1)}
      \frac{\lambda^1}{E_\pi} H_{21}^{(1)} \nn
      & &{} \left.{} +  \frac{1}{2}  H_{21}^{(1)} \frac{\lambda^1}{E_\pi} H_{\pi 
          N} \frac{\lambda^2}{E_\pi} J_{02}^{(-1)} \frac{\lambda^2}{E_\pi} 
        H_{21}^{(1)} \frac{\lambda^1}{E_\pi} H_{21}^{(1)} \right] \eta +
      \text{h.c.}\,.
    \end{eqnarray}
\end{itemize}
%%%%%%%%%%%%%%%%%%%%%%%%%%%%%%%%%%%%%%%%%%%%%%%%%%%%%%%%%%%%%%%%%%%%%%%%%%%%%%%%%

\def\theequation{\Alph{section}.\arabic{equation}}
\setcounter{equation}{0}
\section{Leading two-pion exchange current in momentum space}
\label{app2}
%%%%%%%%%%%%%%%%%%%%%%%%%%%%%%%%%%%%%%%%%%%%%%%%%%%%%%%%%%%%%%%%%%%%%%%%%%%%%%%%%

In this appendix we give the expressions for the two-pion exchange current
operator in momentum space. Following Ref.~\cite{Foldy:1979ie}, the most general
expression for the  current and the charge density can be written as
\beq
\label{Jmom_def}
  \vec{J}  =  \sum_{i=1}^5\sum_{j=1}^{24} f_i^j\left(\vec{q}_1,
    \vec{q}_2\right) \, T_i \vec{O}_j, \qquad J^0 = \sum_{i=1}^5\sum_{j=1}^{8} f_i^{jS}\left(\vec{q}_1,
    \vec{q}_2\right) \, T_i O_j^S \,,
\eeq
where $f_i^j\equiv f_i^j\left(\vec{q}_1 , \vec{q}_2 \right)$ are scalar
functions and the spin-momentum operators $\vec{O}_i$ and $O^S_i$ are given by  
\begin{eqnarray}
  \vec{O}_1    & = & \vec{q}_1+\vec{q}_2,\nn
  \vec{O}_2    & = & \vec{q}_1-\vec{q}_2,\nn
  \vec{O}_3    & = & \cp{q}{\sigma}{1}{2}+\cp{q}{\sigma}{2}{1},\nn
  \vec{O}_4    & = & \cp{q}{\sigma}{1}{2}-\cp{q}{\sigma}{2}{1},\nn
  \vec{O}_5    & = & \cp{q}{\sigma}{1}{1}+\cp{q}{\sigma}{2}{2},\nn
  \vec{O}_6    & = & \cp{q}{\sigma}{1}{1}-\cp{q}{\sigma}{2}{2},\nn
  \vec{O}_7    & = & \vec{q}_1 \spr{q}{q}{\sigma}{1}{2}{2}+\vec{q}_2 \spr{q}{q}{\sigma}{1}{2}{1},\nn
  \vec{O}_8    & = & \vec{q}_1 \spr{q}{q}{\sigma}{1}{2}{2}-\vec{q}_2 \spr{q}{q}{\sigma}{1}{2}{1},\nn
  \vec{O}_9    & = & \vec{q}_2 \spr{q}{q}{\sigma}{1}{2}{2}+\vec{q}_1 \spr{q}{q}{\sigma}{1}{2}{1},\nn
  \vec{O}_{10} & = & \vec{q}_2 \spr{q}{q}{\sigma}{1}{2}{2}-\vec{q}_1 \spr{q}{q}{\sigma}{1}{2}{1},\nn
  \vec{O}_{11} & = & \left(\vec{q}_1+\vec{q}_2 \right)\vek{\sigma}{\sigma}{1}{2},\nn
  \vec{O}_{12} & = & \left(\vec{q}_1-\vec{q}_2 \right)\vek{\sigma}{\sigma}{1}{2},\nn
  \vec{O}_{13} & = & \vec{q}_1\vek{q}{\sigma}{1}{1}\vek{q}{\sigma}{1}{2}+
  \vec{q}_2\vek{q}{\sigma}{2}{1}\vek{q}{\sigma}{2}{2},\nn
  \vec{O}_{14} & = & \vec{q}_1\vek{q}{\sigma}{1}{1}\vek{q}{\sigma}{1}{2}-
  \vec{q}_2\vek{q}{\sigma}{2}{1}\vek{q}{\sigma}{2}{2},\nn
  \vec{O}_{15} & = & \left(\vec{q}_1+\vec{q}_2 \right)\vek{q}{\sigma}{2}{1}\vek{q}{\sigma}{1}{2},\nn
  \vec{O}_{16} & = & \left(\vec{q}_1-\vec{q}_2 \right)\vek{q}{\sigma}{2}{1}\vek{q}{\sigma}{1}{2},\nn
  \vec{O}_{17} & = & \left(\vec{q}_1+\vec{q}_2 \right)\vek{q}{\sigma}{1}{1}\vek{q}{\sigma}{2}{2},\nn
  \vec{O}_{18} & = & \left(\vec{q}_1-\vec{q}_2 \right)\vek{q}{\sigma}{1}{1}\vek{q}{\sigma}{2}{2},\nn
  \vec{O}_{19} & = & \vec{\sigma}_1 \vek{q}{\sigma}{1}{2}+\vec{\sigma}_2\vek{q}{\sigma}{2}{1},\nn
  \vec{O}_{20} & = & \vec{\sigma}_1 \vek{q}{\sigma}{1}{2}-\vec{\sigma}_2\vek{q}{\sigma}{2}{1},\nn
  \vec{O}_{21} & = & \vec{\sigma}_1 \vek{q}{\sigma}{2}{2}+\vec{\sigma}_2\vek{q}{\sigma}{1}{1},\nn
  \vec{O}_{22} & = & \vec{\sigma}_1 \vek{q}{\sigma}{2}{2}-\vec{\sigma}_2\vek{q}{\sigma}{1}{1},\nn
  \vec{O}_{23} & = & \vec{q}_1\vek{q}{\sigma}{2}{1}\vek{q}{\sigma}{2}{2} +
  \vec{q}_2\vek{q}{\sigma}{1}{1}\vek{q}{\sigma}{1}{2},\nn
  \vec{O}_{24} & = & \vec{q}_1\vek{q}{\sigma}{2}{1}\vek{q}{\sigma}{2}{2} -
  \vec{q}_2\vek{q}{\sigma}{1}{1}\vek{q}{\sigma}{1}{2},
\end{eqnarray}
and
\begin{eqnarray}
  O_1^S    & = & \one ,\nn
  O_2^S    & = & \vec q_1 \cdot [ \vec q_2 \times \vec \sigma_2 ] + \vec q_1
  \cdot [ \vec q_2 \times \vec \sigma_1 ]\,, \nn
    O_3^S    & = & \vec q_1 \cdot [ \vec q_2 \times \vec \sigma_2 ] - \vec q_1
  \cdot [ \vec q_2 \times \vec \sigma_1 ]\,, \nn
  O_{4}^S & = & \vec \sigma_1 \cdot \vec \sigma_2 \,, \nn
  O_{5}^S & = & \vek{q}{\sigma}{1}{2}\vek{q}{\sigma}{2}{1},\nn
  O_{6}^S & = & \vek{q}{\sigma}{1}{1}\vek{q}{\sigma}{2}{2},\nn
  O_{7}^S & = &
  \vek{q}{\sigma}{2}{1}\vek{q}{\sigma}{2}{2}+\vek{q}{\sigma}{1}{1}\vek{q}{\sigma}{1}{2},\nn
  O_{8}^S & = &
  \vek{q}{\sigma}{2}{1}\vek{q}{\sigma}{2}{2}-\vek{q}{\sigma}{1}{1}\vek{q}{\sigma}{1}{2}.
\end{eqnarray}
As a basis for the isospin operators we choose
\begin{eqnarray}
  T_1 & = & \tau_1^3 + \tau_2^3,\nn 
  T_2 & = & \tau_1^3 - \tau_2^3,\nn
  T_3 & = & \xp{\tau}{\tau}{1}{2},\nn
  T_4 & = & \vec \tau_1 \cdot \vec \tau_2 \,,\nn
  T_5 & = & \one.
\end{eqnarray}
The nonvanishing long-range contributions to the scalar functions $f_i^j\equiv
f_i^j\left(\vec{q}_1 , \vec{q}_2 \right)$ due to two-pion exchange calculated
using dimensional regularization are given by 
\begin{eqnarray}
\label{fun_f}
  f_3^{1} & = & \frac{i e g_A^2 L(q_1)}{128 \pi ^2 F_\pi^4} \biggl[\frac{g_A^2 (8
      M_\pi^2 + 3 q_1^2)}{4 M_\pi^2 + q_1^2} - 1\biggr]
 + \frac{  e \pi}{F_\pi^4}
\biggl[
     g_A^4 M_\pi^4 \II{d + 2}{2}{1}{2}  + 4 \pi  g_A^4 M_\pi^2 q_1^2
     \II{d + 4}{2}{2}{2} - 8 \pi  g_A^4 M_\pi^2 q_1 \II{d + 4}{3}{1}{2}
(q_1 - q_2
    z) \nn
&&{} - 96 \pi ^2 g_A^4 q_1^3 q_2 z \II{d + 6}{4}{1}{2} + 32 \pi ^2 g_A^4 q_1^2
    q_2 \II{d + 6}{3}{2}{2} (q_1 z + q_2 z^2 + q_2) - 2 \pi  g_A^4 q_1
    \II{d + 4}{2}{1}{2} (q_1 + 2 q_2 z) - 2 (g_A^2 - 1) g_A^2 M_\pi^2 \nn
&&{}\times 
    \II{d + 2}{2}{1}{0} 
- 4 \pi  (g_A^2 - 1) g_A^2 q_1^2 \II{d + 4}{2}{2}{0} + 8 \pi
    (g_A^2 - 1) g_A^2 q_1 \II{d + 4}{3}{1}{0} (q_1 - q_2 z)
 - 2 \pi
    (g_A^2 - 1)^2 \IIR{d + 4}{2}{1}{0}
  \biggr]
 -   (1 \leftrightarrow 2),\nn
  f_3^{2} & = & \frac{i e g_A^2 (g_A^2 + 1)}{256 \pi ^2
    F_\pi^4}
 + \frac{i e g_A^2 L(q_1) }{128 \pi ^2 F_\pi^4}\biggl[\frac{g_A^2 (8 M_\pi^2 + 3 q_1^2)}{4
    M_\pi^2 + q_1^2} - 1\biggr]
 + \frac{e}{8 F_\pi^4}
\biggl[ - 64 \pi ^2 g_A^4 q_1 \II{d + 4}{3}{1}{2}
(M_\pi^2 (q_1 - q_2 z) + q_1^2 q_2 z)\nn
&&{}
 + 16 \pi ^2 g_A^4 q_1
\II{d + 4}{2}{2}{2}  (q_1 q_2^2 (z^2 + 1) - 2 M_\pi^2 (q_1 - q_2 z)) - 2 \pi
g_A^4 \II{d + 2}{1}{1}{2}  (2 M_\pi^2 + q_1 q_2 z)  \nn
&&{}+ 8 \pi  g_A^4 M_\pi^2
\II{d + 2}{2}{1}{2}  (M_\pi^2 + q_1 (q_2 z - q_1)) - 768 \pi ^3 g_A^4 q_1^3
q_2 z \II{d + 6}{4}{1}{2}  - 256 \pi ^3 g_A^4 q_1^2 q_2 \II{d + 6}{3}{2}{2}
(q_1 z - q_2 (z^2 + 1)) \nn
&&{}
 + 16 \pi ^2 g_A^4 q_1 \II{d + 4}{2}{1}{2}  (q_1 - 2
q_2 z) + 2 \pi  (g_A^4 - 1) \IIR{d + 2}{1}{1}{0}  - 8 \pi  (g_A^2 - 1) g_A^2
\II{d + 2}{2}{1}{0} 
(2 M_\pi^2 + q_1 (q_2 z - q_1))  \nn
&&{}+ 32 \pi ^2 (g_A^2 - 1)
g_A^2 q_1 \II{d + 4}{2}{2}{0}  (q_1 - q_2 z) + 64 \pi ^2 (g_A^2 - 1) g_A^2 q_1
\II{d + 4}{3}{1}{0}  (q_1 - q_2 z) - 16 \pi ^2 (g_A^2 - 1)^2
\IIR{d + 4}{2}{1}{0}  \nn
&&{} + g_A^4 M_\pi^4 \II{4}{1}{1}{2}  - 2 (g_A^2 - 1) g_A^2
M_\pi^2 \II{4}{1}{1}{0} \biggr]
 +  (1 \leftrightarrow 2), \nn
f_1^{3} & =& \frac{i e g_A^2 (g_A^2 + 1)}{256 \pi ^2 F_\pi^4}
 + \frac{i e g_A^2 L(q_1)}{128 \pi ^2 F_\pi^4} \biggl[\frac{g_A^2 (8
  M_\pi^2 + 3 q_1^2)}{4 M_\pi^2 + q_1^2} - 1\biggr] + (1 \leftrightarrow 2),
\nn 
f_1^{4} & = & \frac{i e g_A^2 L(q_1)}{128 \pi ^2 F_\pi^4}
\biggl[\frac{g_A^2 (8 M_\pi^2 + 3 q_1^2)}{4 M_\pi^2 + q_1^2} - 1\biggr] -  (1
\leftrightarrow 2), \nn
f_1^{5} & = &  - \frac{i e g_A^4}{128 \pi ^2 F_\pi^4}
 + \frac{i e g_A^4 L(k)}{32 \pi ^2 F_\pi^4}
 - \frac{i e g_A^4 L(q_1)}{64 \pi ^2 F_\pi^4}
 + \frac{  e g_A^2 \pi}{2 F_\pi^4}\biggl[g_A^2 ( - M_\pi^2) \II{d + 2}{1}{1}{2}
 + 4 \pi  g_A^2 q_1 \II{d + 4}{2}{1}{2}  (q_1 - q_2 z)\nn
&&{}
 + (g_A^2 - 1)
\IIR{d + 2}{1}{1}{0} \biggr]
 +  (1 \leftrightarrow 2), \nn
f_1^{6} & = &  - \frac{i e g_A^4 L(q_1)}{64 \pi ^2 F_\pi^4}
 - \frac{2  e g_A^4 \pi ^2}{F_\pi^4} \II{d + 4}{2}{1}{2}  q_1 (q_1 + q_2 z) - 
(1 \leftrightarrow 2), \nn
 f_1^{7} & = & \frac{  e g_A^2 \pi}{2 F_\pi^4}
\biggl[8 \pi  g_A^2 (q_1^2 - 2 M_\pi^2) \II{d + 4}{3}{1}{2}  - g_A^2
M_\pi^2 \II{d + 2}{2}{1}{2}  + 192 \pi ^2 g_A^2 q_1^2 \II{d + 6}{4}{1}{2}
 - 64 \pi ^2 g_A^2 q_1 q_2 z \II{d + 6}{3}{2}{2} 
 + 8 \pi  g_A^2
\II{d + 4}{2}{1}{2} \nn
&&{}  + (g_A^2 - 1) \II{d + 2}{2}{1}{0}  + 16 \pi  (g_A^2 - 1)
\II{d + 4}{3}{1}{0} \biggr]
 -  (1 \leftrightarrow 2),\nn
f_1^{8} & = & \frac{i e g_A^4}{32 \pi ^2 F_\pi^4 k^2}
 - \frac{i e g_A^4 L(k)}{32 \pi ^2 F_\pi^4 k^2}
 + \frac{ e g_A^2\pi }{4 F_\pi^4}
\biggl[16 \pi  g_A^2 (q_1^2 - 2 M_\pi^2) \II{d + 4}{3}{1}{2}  - 2 g_A^2
M_\pi^2 \II{d + 2}{2}{1}{2}  + 384 \pi ^2 g_A^2 q_1^2 \II{d + 6}{4}{1}{2} \nn
&&{}
 - 8
\pi  g_A^2 q_1 q_2 z \II{d + 4}{2}{2}{2}  - 128 \pi ^2 g_A^2 q_1 q_2 z
\II{d + 6}{3}{2}{2}  + g_A^2 \II{d + 2}{1}{1}{2}  + 16 \pi  g_A^2
\II{d + 4}{2}{1}{2}  + 2 (g_A^2 - 1) \II{d + 2}{2}{1}{0} \nn
&&{}
 + 32 \pi
(g_A^2 - 1) \II{d + 4}{3}{1}{0} \biggr]
 +  (1 \leftrightarrow 2),\nn
  f_1^{9} & =& \frac{ e g_A^2\pi }{2 F_\pi^4}
\biggl[g_A^2 M_\pi^2 \II{d + 2}{2}{1}{2}  - 8 \pi  g_A^2 q_1^2
\II{d + 4}{3}{1}{2}  - 64 \pi ^2 g_A^2 q_1 \II{d + 6}{3}{2}{2}  (q_1 + q_2
z) + 4 \pi  g_A^2 \II{d + 4}{2}{1}{2}  + (1 - g_A^2) \II{d + 2}{2}{1}{0}
\biggr]\nn
&&{}  -  (1 \leftrightarrow 2), \nn
  f_1^{10}& = & \frac{i e g_A^4}{32 \pi ^2 F_\pi^4 k^2}
 - \frac{i e g_A^4 L(k)}{32 \pi ^2 F_\pi^4 k^2}
 - \frac{  e g_A^2\pi}{4 F_\pi^4}
\biggl[ - 8 \pi  g_A^2 \II{d + 4}{2}{2}{2}  (2 M_\pi^2 + q_1 q_2 z) - 2 g_A^2
M_\pi^2 \II{d + 2}{2}{1}{2}  + 16 \pi  g_A^2 q_1^2 \II{d + 4}{3}{1}{2} \nn
&&{}
 + 128
\pi ^2 g_A^2 q_1 \II{d + 6}{3}{2}{2}  (q_1 - q_2 z) + g_A^2
\II{d + 2}{1}{1}{2}  + 8 \pi  g_A^2 \II{d + 4}{2}{1}{2}  + 2 (g_A^2 - 1)
\II{d + 2}{2}{1}{0}  + 16 \pi  (g_A^2 - 
1) \II{d + 4}{2}{2}{0} \biggr]\nn
&&{}
 +  (1 \leftrightarrow 2)\,.
\end{eqnarray}
In addition, there are nonvanishing  functions $f_2^j$ given by 
\beq
  f_2^3   = f_1^4, \;\;\; f_2^4 = f_1^3, \;\;\; f_2^5 = -f_1^6, \;\;\;
  f_2^6 = - f_1^5, \;\;\; f_2^7 = f_1^8, \;\;\; f_2^8 = f_1^7, \;\;\; f_2^9 = f_1^{10}, \;\;\;
  f_2^{10} = f_1^9.
\eeq
In the above equations, $z \equiv \hat{q}_1 \cdot\hat{q}_2$, $q_i \equiv \left|\vec{q_i}
\right|$ and the loop functions $L(q)$ and $A(q)$ are defined in Eq.~(\ref{LundA}). 
Further, the functions $I$ correspond to the three-point functions via 
\beq
  \II{d}{\nu_1}{\nu_2}{\nu_3}  \equiv I(d;0,1;q_1,\nu_1; - q_2,\nu_2; 0,\nu_3 )
  \quad \textrm{with} \ \; \; q_{i}  = (0,\vec q_{i})
\eeq
and 
\beq
\label{def3point}
I(d;p_1,\nu_1;p_2,\nu_2;p_3,\nu_3;p_4,\nu_4)  = \mu^{4-d}\int \frac{d^d\ell}{\left(
    2\pi\right)^d} \frac{1}{[ \left(\ell+p_1\right)^2 - M_\pi^2 ]^{\nu_1} \,[ ( \ell +
    p_2)^2-M_\pi^2]^{\nu_2} [ ( \ell +
    p_3 )^2 -M_\pi^2 ]^{\nu_3} [ v \cdot \left(\ell + p_4 \right) ]^{\nu_4}} \,.
\eeq
Here, all propagators are understood to have an infinitesimal  positive
imaginary part. Notice that here and in what follows, we will only need the functions $I$ for
four-momenta with vanishing zeroth component.
We further emphasize that all functions $\II{d+n}{\nu_1}{\nu_2}{\nu_3}$ which enter the above
equations except $\II{d+2}{1}{1}{0}$, $\II{d+4}{2}{1}{0}$ and
$\II{d+4}{1}{2}{0}$ are finite in dimensional regularization in the limit $d\rightarrow
4$. For these
functions, we define reduced functions by subtracting the poles in four dimensions
\begin{eqnarray}
  \IIR{d+2}{1}{1}{0} & = & \II{d+2}{1}{1}{0} - \frac{i}{4\pi}L(\mu) -
  \frac{i}{128\pi^3}\ln\left( \frac{M_\pi^2}{\mu^2}\right),\nn
  \IIR{d+4}{2}{1}{0} & = & \II{d+4}{2}{1}{0} + \frac{i}{48\pi^2}L(\mu) +
  \frac{i}{1536\pi^4}\ln\left( \frac{M_\pi^2}{\mu^2}\right),\nn
 \IIR{d+4}{1}{2}{0} & = & \II{d+4}{1}{2}{0} + \frac{i}{48\pi^2}L(\mu) +
  \frac{i}{1536\pi^4}\ln\left( \frac{M_\pi^2}{\mu^2}\right) \,,
\end{eqnarray}
where 
\beq
L(\mu) =  \frac{\mu^{d-4}}{16 \pi^2}\left[ \frac{1}{d-4} +
\frac{1}{2}\left( \gamma_{\rm E} - 1 - \ln \left(4\pi \right)\right)\right].
\eeq
Here, $\mu$ is the scale introduced in dimensional regularization and
$\gamma_{\rm E} = - \Gamma^\prime (1) \simeq 0.577$.
Finally, for scalar functions contributing to the charge density we obtain the
following expressions:
\begin{eqnarray}
  f_3^{2S} & =& \frac{e g_A^4 A(k)}{64 \pi
    F_\pi^4}+\frac{i \pi ^2 e g_A^4}{F_\pi^4}\biggl[-2 M_\pi^2
  \II{d+4}{2}{1}{3} +16 \pi  q_1^2 \II{d+6}{3}{1}{3} -8 \pi  q_1 q_2 z
  \II{d+6}{2}{2}{3} +\II{d+4}{1}{1}{3} \biggr]+ (1 \leftrightarrow 2),
\nn
  f_3^{3S} & = & \frac{2 i \pi ^2 e g_A^4}{F_\pi^4}\biggl[8
  \pi  q_1^2 \II{d+6}{3}{1}{3} -M_\pi^2 \II{d+4}{2}{1}{3} \biggr]- (1
  \leftrightarrow 2),\nn
  f_1^{1S} & = & \frac{e g_A^4 M_\pi (12 M_\pi^4+7 M_\pi^2
    q_1^2+q_1^2 q_2^2)}{64 \pi  F_\pi^4 (4 M_\pi^2+q_1^2) (4
    M_\pi^2+q_2^2)}
-\frac{e g_A^4 A(k) (2 M_\pi^2+q_1^2)}{16 \pi  F_\pi^4}
+\frac{e g_A^4 A(q_1) (2 M_\pi^2+q_1^2)}{32 \pi  F_\pi^4}\nn
&&{}
+\frac{i   e g_A^4\pi}{F_\pi^4}
 \biggl[M_\pi^4 (-\II{d+2}{1}{1}{3} )+8 \pi  M_\pi^2 q_1
\II{d+4}{2}{1}{3}  (q_1-q_2 z)+64 \pi ^2 q_1^3 q_2 z \II{d+6}{3}{1}{3}
-16 \pi ^2 q_1^2 q_2^2 (z^2+1) \II{d+6}{2}{2}{3} \nn
&&{}
+2 \pi  q_1 q_2 z
\II{d+4}{1}{1}{3} \biggr]
+ (1 \leftrightarrow 2), \nn
  f_1^{4S} & = & -\frac{e g_A^4 q_1^2 A(q_1)}{64 \pi
    F_\pi^4}
-\frac{2 i  e g_A^4\pi ^2}{F_\pi^4}\biggl[8 \pi  q_1^2 q_2^2 (z^2-1)
\II{d+6}{2}{2}{3} -q_1 q_2 z \II{d+4}{1}{1}{3} \biggr]
+ (1 \leftrightarrow 2),\nn
  f_1^{5S} & = & \frac{2 i  e g_A^4\pi^2}{F_\pi^4}
\biggl[8 \pi  q_1 q_2 z \II{d+6}{2}{2}{3} -\II{d+4}{1}{1}{3} \biggr]+
(1 \leftrightarrow 2),\nn
  f_1^{6S}  &=&  \frac{16 i \pi ^3 e g_A^4}{F_\pi^4}
q_1 q_2 z \II{d+6}{2}{2}{3} + (1 \leftrightarrow 2),\nn
  f_1^{7S} & = & \frac{e g_A^4 A(q_1)}{128 \pi  F_\pi^4}
-\frac{16 i \pi ^3 e g_A^4}{F_\pi^4}
q_1^2 \II{d+6}{2}{2}{3} + (1 \leftrightarrow 2), \nn
  f_1^{8S} &=&  -\frac{e g_A^4 A(q_1)}{128 \pi  F_\pi^4}
-\frac{16 i \pi ^3 e g_A^4}{F_\pi^4}q_1^2 \II{d+6}{2}{2}{3} 
- (1 \leftrightarrow 2), \nn
  f_2^{1S} & = & \frac{e g_A^4 M_\pi^3 q_1^2}{64 \pi  F_\pi^4
    (4 M_\pi^2+q_1^2) (4 M_\pi^2+q_2^2)}
+\frac{e g_A^2 (g_A^2-1) A(q_1) (2 M_\pi^2+q_1^2)}{32 \pi  F_\pi^4}- (1 \leftrightarrow 2),
\nn
  f_2^{7S} & = & -\frac{e g_A^4 A(q_1)}{128 \pi  F_\pi^4}-
  (1 \leftrightarrow 2), \nn
 f_2^{8S} &=&  \frac{e g_A^4 A(q_1)}{128 \pi  F_\pi^4}+ (1
  \leftrightarrow 2).
\end{eqnarray}

\def\theequation{\Alph{section}.\arabic{equation}}
\setcounter{equation}{0}
\section{Evaluation of the 3-point function}
\label{app3}
%%%%%%%%%%%%%%%%%%%%%%%%%%%%%%%%%%%%%%%%%%%%%%%%%%%%%%%%%%%%%%%%%%%%%%%%%%%%%%%%%

The momentum-space expressions given in the previous appendix involve the
3-point function defined in Eq.~(\ref{def3point}). Below we show how
loop integrals of this kind  can be evaluated by introducing the corresponding Feynman
parameters. In particular, we consider the following integrals 
\beqa
\label{Is_def}
I\!\!I_1 &=& I(d;0,\nu_1;p_2,\nu_2;p_3,\nu_3;0,\nu_4)\,, \nn 
I\!\!I_2 &=& I(d;0,\nu_1;p_2,\nu_2;p_3,\nu_3;0,0)\,,
\eeqa 
in the following kinematics:
\beq
  q_i  =  (0,\, \vec{q}_i),  \ \ \ 
  \hat{q}_1\cdot\hat{q}_2=z, \ \ \   v^2 =  1, \ \ \  v\cdot q_1 = v\cdot q_2 = 0. 
\eeq
Here, we consider the case with $\nu_i = 1, 2, 3, \ldots$ and $|\vec{q}_i \, |
> 0$. 
Starting from this point, we denote by  $q_i$ the length of the corresponding
three-momentum in units of the pion mass, i.e.~ $q_i \equiv |\vec{q}_i \, | / M_\pi$. 
Introducing the Feynman parameters and carrying out the integration over the
loop momentum, we obtain the following result for the first integral in Eq.~(\ref{Is_def}):
\beqa
\label{I1final}
I\!\!I_1 &=& \mu^{4-d}\frac{2^{\nu_4-1}\Gamma (\frac{\nu_4}{2})\Gamma\left(\nu_1+\nu_2+\nu_3+\frac{\nu_4}{2} - \frac{d}{2}
  \right)}{\Gamma\left(\nu_1\right)\Gamma\left(\nu_2\right)\Gamma\left(\nu_3
  \right)\Gamma\left(\nu_4  \right)}\frac{(-1)^{\nu_1+ \nu_2 +
    \nu_3 + \nu_4} i }{(4\pi)^{(d/2)}}
\int_0^1 dt \int_0^{t} dy  \left(t-y
  \right)^{\nu_1-1} \, \left(1-t \right)^{\nu_2-1} \,  y^{\nu_3-1} \nn
&\times &\left(\frac{1}{M_\pi^2D (y_1 - y_2)}\right)^{
    \nu_1+ \nu_2 + \nu_3 +  \frac{\nu_4}{2} - \frac{d}{2}} 
 \left(\frac{1}{y_1 - y} + \frac{1}{y -y_2 } \right)^{
    \nu_1+ \nu_2 + \nu_3 +  \frac{\nu_4}{2} - \frac{d}{2}}\,,
\eeqa
where 
\beq
y_1  =  \frac{E}{2D}+\sqrt{\frac{E^2+4AD}{4D^2}} \,, \quad \quad
y_2  =  \frac{E}{2D}-\sqrt{\frac{E^2+4AD}{4D^2}}\,,
\eeq
and we have introduced 
\beqa
A & \equiv & 1+ q_1^2(1-t)t \;  >\;  0\,,\nn
B & \equiv & 2 q_1 q_2 z + q_2^2\,, \nn
C& \equiv & -2 q_1q_2z\,, \nn  
D& \equiv &q_2^2 \; >\;  0\,, \nn
E & \equiv & B +tC \,.
\eeqa
It can be shown that the following inequalities hold true in the integration
region for the considered
kinematics:
\beq
\label{ineq}
  y_1 > t \geq y \geq 0 > y_2\,.
\eeq
Thus, the remaining  two-dimensional integral in Eq.~(\ref{I1final}) can be
easily calculated numerically for all desired values of $\nu_i$. 

Similarly, for the second integral in Eq.~(\ref{Is_def}) we obtain:
\beqa
\label{I2final}
I\!\!I_2 &=& 
\mu^{4-d}\frac{\Gamma\left(\nu_1+\nu_2+\nu_3- \frac{d}{2}
  \right)}{\Gamma\left(\nu_1\right)\Gamma\left(\nu_2\right)\Gamma\left(\nu_3
  \right) }\frac{(-1)^{\nu_1+ \nu_2 +
    \nu_3 } i }{(4\pi)^{(d/2)}} 
\int_0^1 dt \int_0^{t} dy  \left(t-y
  \right)^{\nu_1-1} \, \left(1-t \right)^{\nu_2-1} \,  y^{\nu_3-1} \nn
&\times& 
\left(\frac{1}{M_\pi^2D (y_1 - y_2)}\right)^{
    \nu_1+ \nu_2 + \nu_3  - \frac{d}{2}} 
 \left(\frac{1}{y_1 - y} + \frac{1}{y -y_2 } \right)^{
    \nu_1+ \nu_2 + \nu_3   - \frac{d}{2}} \,,
\eeqa
where, again, the inequalities given in Eq.~(\ref{ineq}) hold true. 

For the reduced functions $\IIR{d+2}{1}{1}{0}$, $\IIR{d+4}{1}{2}{0}$ and $\IIR{d+4}{2}{1}{0}$
 we obtain the expressions:

\beqa
\IIR{d+2}{1}{1}{0} & = & \frac{i}{64 \pi^3}\int_0^1 dt \int_0^t dy \left(1 -
  \ln\left(\frac{1}{D(y_1-y)(y-y_2)} \right) \right)\, ,\nn
\IIR{d+4}{2}{1}{0} & = & \frac{i}{256 \pi^4}\int_0^1 dt \int_0^t dy \, (t-1)\left(1-
  \ln\left(\frac{1}{D(y_1-y)(y-y_2)} \right) \right)\, ,\nn
\IIR{d+4}{1}{2}{0} & = & -\frac{i}{256 \pi^4}\int_0^1 dt \int_0^t dy \, y \left(1-
  \ln\left(\frac{1}{D(y_1-y)(y-y_2)} \right) \right)\, ,
\eeqa
that do not depend on $\mu$.

\def\theequation{\Alph{section}.\arabic{equation}}
\setcounter{equation}{0}
\section{Current conservation and the continuity equation}
\label{app4}
%%%%%%%%%%%%%%%%%%%%%%%%%%%%%%%%%%%%%%%%%%%%%%%%%%%%%%%%%%%%%%%%%%%%%%%%%%%%%%%%%

Current conservation implies that the electromagnetic current operator $\vec{J}(\vec{x})$ should
fulfill the continuity equation 
\beq
\vec{\nabla}\cdot \vec{J}(\vec{r}) = - \frac{\partial \rho}{\partial t} = -i
  \left[H,\rho \right] = -i\left[H_0+ \bar V,\rho \right]\,,
\eeq
where $\rho = \rho (\vec{r})  \equiv J^0(\vec{r})$ is the charge density and
$H$, $H_0$ and $\bar V$ refer to 
the two-nucleon Hamilton operator, kinetic energy term and the potential,
respectively. The continuity equation thus provides a powerful check for the
calculation. For the leading two-pion exchange contributions to the current
operator the continuity equation takes the form  
\beq
\label{cont_2pi_coord}
\vec{\nabla}\cdot\vec{J}_{2\pi}^{\,(1)}(\vec{r}) = e \xp{\tau}{\tau}{1}{2}
W_{2\pi}^{(2)}(\vec{r}_1-\vec{r}_2)
\left[\delta (\vec{r}-\vec{r}_1)-\delta (\vec{r}-\vec{r}_2)\right], 
\eeq
where the potential in the isospin limit is written in the form 
\beq
\bar V = V + \vec \tau_1 \cdot \vec \tau_2 \, W\,,
\eeq
$\vec r_i$ denotes the position of the nucleon $i$ 
and the superscripts of $\vec{J}$ and $W$ refer to the chiral order. Further,
we made use of the explicit form of the leading one-body charge density 
\beq
\rho (\vec{r})= \frac{e}{2} \left[  \left(\one +
    \tau_1^3 \right) \delta (\vec{r}-\vec{r}_1) + \left(\one +
    \tau_2^3 \right) \delta (\vec{r}-\vec{r}_2) \right]\,,
\eeq
and the absence of the leading one-pion exchange charge density,
$\rho_{1\pi}^{(-1)} = 0$. Eq.~(\ref{cont_2pi_coord}) 
can be transformed into momentum space 
\beq
\label{cont_mom}
i (\vec{q}_1+\vec{q}_2)\cdot \vec{J}_{2\pi}^{\,(1)}(\vec{q}_1,\vec{q}_2) = e
  \xp{\tau}{\tau}{1}{2} \left[W_{2\pi}^{(2)}(\vec{q}_1)- W_{2\pi}^{(2)}(\vec{q}_2)\right]\,.
\eeq
The left-hand side of the above equation can be expressed in terms of the
basis operators $O_i^S$ defined in Appendix~\ref{app2}. Using the
representation of the current operator in Eq.~(\ref{Jmom_def}), 
the spin-momentum operators appearing on the left-hand side of the above
equation can be expressed in terms of the operators $O_j^S$ as follows:
\beq
\sum_{j=1}^{24} f_i^j\left(\vec{q}_1,
    \vec{q}_2\right) \, \vec{k}\cdot\vec{O}_j = \sum_{j=1}^{8} g_i^{jS}\left(\vec{q}_1,
    \vec{q}_2\right) \, O_j^S \,,
\eeq
where the scalar functions $g_i^{jS}\left(\vec{q}_1, \vec{q}_2\right)$
read
\beqa
  g^{1S}_{i} & = & k^2 f_{i}^1 + (q_1^2 - q_2^2 ) f_{i}^2,\nn
  g^{2S}_{i} & = & -f_{i}^4 - f_{i}^6+\frac{1}{2}k^2 f_{i}^7 +
  \frac{1}{2} (q_1^2 - q_2^2 ) f_{i}^8+\frac{1}{2}k^2 f_{i}^9 -
  \frac{1}{2} (q_1^2 - q_2^2 )  f_{i}^{10},\nn
  g^{3S}_{i} & = & -f_{i}^3 + f_{i}^5+\frac{1}{2}k^2 f_{i}^8 +
  \frac{1}{2} (q_1^2 - q_2^2 )  f_{i}^7+\frac{1}{2}k^2 f_{i}^{10} -
  \frac{1}{2} (q_1^2 - q_2^2 )  f_{i}^9,\nn
  g^{4S}_{i} & = & k^2 f_{i}^{11} +  (q_1^2 - q_2^2 )  f_{i}^{12},\nn
  g^{5S}_{i} & = & k^2 f_{i}^{15} + (q_1^2 - q_2^2 )  f_{i}^{16} + 2f_i^{19},\nn
  g^{6S}_{i} & = & k^2 f_{i}^{17} + (q_1^2 - q_2^2 )  f_{i}^{18} + 2f_i^{21},\nn
  g^{7S}_{i} & = & f_{i}^{19} + f_{i}^{21}+\frac{1}{2}k^2 f_{i}^{13} +
  \frac{1}{2} (q_1^2 - q_2^2 )  f_{i}^{14}+\frac{1}{2}k^2 f_{i}^{23} +
  \frac{1}{2} (q_1^2 - q_2^2 )  f_{i}^{24},\nn
  g^{8S}_{i} & = & - f_{i}^{20} + f_{i}^{22}-\frac{1}{2}k^2 f_{i}^{14} -
  \frac{1}{2} (q_1^2 - q_2^2 )   f_{i}^{13}+\frac{1}{2}k^2 f_{i}^{24} +
  \frac{1}{2} (q_1^2 - q_2^2 )  f_{i}^{23}\,,
\eeqa
and $\vec k \equiv \vec{q}_1+\vec{q}_2 $. 
Substituting the expressions for $f_i^j$ for the leading two-pion exchange
current operator from Eq.~(\ref{fun_f}), one finds that the
only non-vanishing function $g_i^{jS\, (2\pi)}$ is given by 
\beq
  g_{3}^{1S \, (2\pi)}  =   e \frac{i L(q_1)}{384 \pi^2 F_\pi^4}
\biggl\{4M_\pi^2\left(5g_A^4-4g_A^2-1 \right) + q_1^2\left(23g_A^4
    -10g_A^2-1\right) + \frac{48 g_A^4M_\pi^4}{q_1^2+4M_\pi^2}\biggr\}\nn
%   e \frac{i
%     q_1^2}{384F_\pi^4\pi^2}\left(23g_A^4-10g_A^2-1
% \right)\log\left( \frac{M_\pi}{\mu}\right) 
+ \alpha
-(1\leftrightarrow 2)\,.
\eeq
 where $\alpha$ denotes a polynomial in $q_1$ and $q_2$ whose form depends on the
 choice of the subtraction scheme.
Using the explicit form of $W_{2\pi}^{(2)}$, 
\beq
  W_{2\pi}^{(2)}  =   -\frac{L(q)}{384 \pi^2F_\pi^4} 
  \biggl\{4M_\pi^2\left(5 g_A^4 - 4 g_A^2 -1 \right)  + q^2 \left( 23 g_A^4 -
    10 g_A^2 -1\right) + \frac{48 g_A^4M_\pi^4}{4M_\pi^2+q^2}\biggr\}
%\nn
% &+& \frac{1}{384\pi^2F_\pi^4} \bigg\{-18M_\pi^2
% \left(5g_A^4-2g_A^2 \right)\log\left(\frac{M_\pi}{\mu}\right)
% + q^2 \left(-23g_A^4+10g_A^2 +1
% \right)\log\left(\frac{M_\pi}{\mu}\right)\biggr\} 
+ \alpha^\prime
\,,
\eeq
where $\alpha^\prime$ again denotes a subtraction-scheme dependent polynomial contribution,
it is easy to see that Eq.~(\ref{cont_mom}) is indeed fulfilled.

\def\theequation{\Alph{section}.\arabic{equation}}
\setcounter{equation}{0}
\section{Coordinate-space representation of various loop integrals}
\label{app5}
%%%%%%%%%%%%%%%%%%%%%%%%%%%%%%%%%%%%%%%%%%%%%%%%%%%%%%%%%%%%%%%%%%%%%%%%%%%%%%%%%

In this appendix we collect the formulae needed to obtain the expressions for
various loop integrals in coordinate-space. 
The two-pion exchange contributions to the current operator resulting in the
method of unitary transformation are given in terms of three-dimensional loop
integrals with the integrands being rational functions of the pion energies 
$\omega_i = \sqrt{k_i^2 + M_\pi^2}$ where $k_i \equiv | k_i \, |$ refers to
the pion momentum. In order to end up with simple expressions in coordinate
space, it is convenient to rewrite the integrands as continuous superpositions
of the propagators using the following expressions:
\begin{eqnarray}
\label{omegas}
  \frac{1}{\omega_1\omega_2(\omega_1+\omega_2)}&=&\frac{2}{\pi}\int_0^\infty
  d\beta \frac{1}{\omega_1^2+\beta^2} \frac{1}{\omega_2^2+\beta^2}\,,\nn
\frac{\omega_1^2+\omega_1\omega_2+\omega_2^2}{\omega_1^3\omega_2^3(\omega_1+\omega_2)}&=&{}-\frac12
\frac{\partial}{\partial
M_\pi^2}\frac{1}{\omega_1\omega_2(\omega_1+\omega_2)}\,,\nn
 & = & \frac{1}{\pi}\int_0^\infty d\beta
 \biggl(\frac{1}{(\omega_1^2+\beta^2)^2(\omega_2^2+\beta^2)}+\frac{1}{(\omega_1^2+\beta^2)
   (\omega_2^2+\beta^2)^2}   
 \biggr) \,.
\end{eqnarray}
In addition, we also need the following relations which involve three
different pion energies:
\beqa
\label{3omegas}
D_1 &\equiv& =\frac{1}{\omega_1\omega_2\omega_3} 
\frac{\omega_1+\omega_2+\omega_3}{(\omega_1+\omega_2)(\omega_1+\omega_3)(\omega_2+\omega_3)}
\nn
&=& \frac{2}{\pi}\int_0^\infty
d\beta\frac{1}{(\omega_1^2+\beta^2)(\omega_2^2+\beta^2)(\omega_3^2+\beta^2)}\,, \nn
D_2&\equiv& \frac{1}{\omega_1\omega_2\omega_3}\left(\frac{(\omega_1-\omega_2)
    (\omega_1-\omega_3)}{(\omega_1+\omega_2) 
   (\omega_1+\omega_3)}  +  \frac{(\omega_1+\omega_2)
   (\omega_1-\omega_3)}{(\omega_1+\omega_3)
   (\omega_2+\omega_3)}+\frac{(\omega_1-\omega_2)
   (\omega_1+\omega_3)}{(\omega_1+\omega_2) (\omega_2+\omega_3)}\right) \nn
&=& -\frac{4}{\pi}\int_0^\infty d\beta \left(\frac{4\,
    \beta^2}{(\omega_1^2+\beta^2)(\omega_2^2+\beta^2)(\omega_3^2+\beta^2)}-\frac{1}{
    (\omega_2^2+\beta^2)(\omega_3^2+\beta^2)}\right) \,, \nn
D_3& \equiv&  \frac{1}{\omega_1\omega_2\omega_3} \left(
    \frac{1}{\omega_1\omega_2\omega_3^2}-\frac{\omega_2}{\omega_1^2(\omega_1^2-\omega_2^2)
      (\omega_1+\omega_3)}+\frac{\omega_1}{\omega_2^2(\omega_1^2-\omega_2^2)(\omega_2+\omega_3)}
  \right)\nn
&=&  -2\frac{\partial}{\partial M_\pi^2} \frac{1}{\omega_1\omega_2\omega_3}
\frac{\omega_1+\omega_2+\omega_3}{(\omega_1+\omega_2)(\omega_1+\omega_3)(\omega_2+\omega_3)}\\
&=& \frac{4}{\pi}\int_0^\infty d\beta \biggl[
 \frac{1}{(\omega_1^2+\beta^2)^2(\omega_2^2+\beta^2)(\omega_3^2+\beta^2)}+\frac{1}{(\omega_1^2+\beta^2)
   (\omega_2^2+\beta^2)^2(\omega_3^2+\beta^2)}
+\frac{1}{(\omega_1^2+\beta^2)(\omega_2^2+\beta^2)(\omega_3^2+\beta^2)^2}\biggr]\,
.\nonumber
\eeqa
Using the above expressions, it is straightforward to carry out the
Fourier transformation $\mathcal{F}$, 
\beq
\mathcal{F} \left( f (\vec q_1 \,, \, \vec q_2 \,) \right) \equiv
\int \frac{d^3q_1}{(2\pi)^3} \frac{d^3q_2}{(2\pi)^3} e^{i
  \vec{q}_1 \cdot \vec{r}_{10}}  e^{i  \vec{q}_2 \cdot \vec{r}_{20}}\, f (\vec
q_1 \,, \, \vec q_2 \,) \,,
\eeq
of the integrals which appear in the calculation. For example,
\beqa
\mathcal{F} \left(\int
\frac{d^3 \ell}{(2\pi)^3}\frac{1}{\omega_+\omega_-(\omega_++\omega_-)}  \right)
 & = &\delta (\vec r_{20}) \int \frac{d^3q_1}{(2\pi)^3} \int
\frac{d^3 \ell}{(2\pi)^3} e^{i  \vec{q}_1 \cdot
   \vec{r}_{10}} \frac{2}{\pi} \int_0^\infty d\beta
 \frac{1}{(\omega_+^2+\beta^2)(\omega_-^2+\beta^2)}\nn
& = &\delta (\vec r_{20})\frac{16}{\pi} \int_0^\infty d\beta \int
\frac{d^3k_1}{(2\pi)^3} \frac{d^3k_2}{(2\pi)^3}e^{i  \vec{k}_1 \cdot 
   \vec{r}_{10}}e^{-i  \vec{k}_2
   \cdot\vec{r}_{10}}\frac{1}{(4\omega_1^2+\beta^2)(4\omega_2^2+\beta^2)} \nn
&=& \frac{ M_\pi^3}{4 \pi^3 } \,  \delta (\vec r_{20}) \, \int_0^\infty d\beta
 \frac{e^{-\sqrt{1+\beta^2} \,2  x_{10}}}{2 x^2_{10}} \nn
&=& \frac{M_\pi^6}{8 \pi^3}\,  \delta (\vec x_{20})\, \frac{K_1
   (2 x_{10})}{x_{10}^2}\,,
\eeqa
where $\omega_\pm = \sqrt{(\vec l \pm \vec q_1)^2 + 4 M_\pi^2}$ and the $K_i$
refer to the modified Bessel functions. 
We made use of Eq.~(\ref{omegas}) in the first line  and 
carried out the substitutions $\vec{\ell}+\vec{q}_1 \rightarrow 2
\vec{k}_1$, $\vec{\ell}-\vec{q}_1  \rightarrow 2 \vec{k}_2$ in the second line. Notice further
that the last equality is valid for $x_{10} > 0$. In a similar way one obtains
\beqa
\mathcal{F} \left(\int
\frac{d^3 \ell}{(2\pi)^3}\frac{l^2}{\omega_+\omega_-(\omega_++\omega_-)}  \right)
 & = & \frac{M_\pi^8}{12 \pi^3}\,  \delta (\vec x_{20})\, \left( \nabla_{10}^2
   - 4 \right) \frac{K_1
   (2 x_{10})}{x_{10}^2}\,, \nn
\mathcal{F} \left( \int
\frac{d^3 \ell}{(2\pi)^3}\frac{\omega_+^2 + \omega_+ \omega_- +
  \omega_-^2}{\omega_+^3\omega_-^3(\omega_++\omega_-)}  \right)
 & = & \frac{M_\pi^4}{16 \pi^3} \delta (\vec x_{20})
 \frac{K_0(2x_{10})}{x_{10}}  \,. 
\eeqa
Again, these expressions are only valid for $x_{10} > 0$. Clearly, the above
results may also be obtained by first carrying out the momentum-space loop
integrals. The resulting non-polynomial expressions are given in terms of the
loop functions $L$ and $A$, see Eq.~(\ref{LundA}), and can be easily
Fourier-transformed using the following relations 
\beqa 
  \int \frac{d^3q}{(2\pi)^3} e^{i \vec{q}\cdot \vec{r}}\frac{L(q)}{s^2} &
  \rightarrow & \frac{M_\pi}{4 \pi} \frac{K_0(2x)}{x}, \nn
  \int \frac{d^3q}{(2\pi)^3} e^{i \vec{q}\cdot \vec{r}} (s^2)^k L(q) &
  \rightarrow & (-1)^{(k+1)} (\nabla_x^2 - 4)^k\frac{M_\pi^{(3+2k)}}{2 \pi} \frac{K_1(2x)}{
    x^2}\,, \nn
  \int \frac{d^3q}{(2\pi)^3} e^{i \vec{q}\cdot \vec{r}} (s^2)^k A(q) &
  \rightarrow & (-1)^k(\nabla_x^2 - 4)^k\frac{M_\pi^{2+2k}}{8\pi} \frac{e^{-2 x}}{x^2} \,,
\eeqa
where $k=0,1,2,\ldots$.

Consider now the more complicated expressions involving three different pion
energies which appear in the calculation:
\beqa
&& \mathcal{F} \left( \int\frac{d^3 k_1}{(2\pi)^3}d^3 k_2 d^3 k_3\,
\delta (\vec k_2-\vec k_1+\vec q_1) \, \delta ( \vec k_3- \vec k_1- \vec q_2) \, (\vec k_2 + \vec k_3 ) \,
D_2 \right) \nn
&=& {}-\frac{4}{\pi}\int_0^\infty d\beta \int \frac{d^3 k_1}{(2\pi)^3}\frac{d^3 k_2}{(2\pi)^3}\frac{d^3 k_3}{(2\pi)^3}e^{-i\vec{k}_2\cdot\vec{r}_{10}}e^{i\vec{k}_3\cdot\vec{r}_{20}}e^{i\vec{k}_1\cdot\vec{r}_{12}}\biggl(\frac{4\,
  \beta^2}{(\omega_1^2+\beta^2)(\omega_2^2+\beta^2)(\omega_3^2+\beta^2)}
-\frac{1}{(\omega_2^2+\beta^2)(\omega_3^2+\beta^2)}
\biggr)\nn
&& {} \times (\vec{k}_2+\vec{k}_3) \nn
&=& \frac{iM_\pi^7}{4 \pi^4}(\vec{\nabla}_{10}-\vec{\nabla}_{20})\int_0^\infty  d
\beta \beta^2 \, 
 \frac{e^{-\sqrt{1+\beta^2} \, x_{10}}}{x_{10}}  \,
 \frac{e^{-\sqrt{1+\beta^2} \, x_{20}}}{x_{20}}   \,
 \frac{e^{-\sqrt{1+\beta^2} \, x_{12}}}{x_{12}}  \nn
&=& \frac{i M_\pi^7}{4 \pi^4} (\vec{\nabla}_{10} -
  \vec{\nabla}_{20}) \frac{K_2(x_{10} + x_{20} +
    x_{12})}{(x_{10}\, x_{20}\, x_{12})(x_{10} + x_{20} + x_{12})}\,,
\eeqa
where in the first line we used Eq.~(\ref{3omegas}). The last relation is 
valid for positive values of the argument of the Bessel function.  Similarly,
we obtain:
\beqa
&& \mathcal{F} \left( \int\frac{d^3 k_1}{(2\pi)^3}d^3 k_2 d^3 k_3\,
\delta (\vec k_2-\vec k_1+\vec q_1) \, \delta ( \vec k_3- \vec k_1- \vec
q_2) \, k_1^{i_1}\cdots k_1^{i_m} k_2^{j_1}\cdots
  k_2^{j_n} k_3^{k_1} \cdots k_3^{k_p} \, D_1 \right) \nn
&=& (-1)^{m+p} \frac{M_\pi^4 (i M_\pi )^{n+m+p}}{32\pi^4}
\nabla_{12}^{i_1}\cdots\nabla_{12}^{i_m} \nabla_{10}^{j_1}\cdots 
  \nabla_{10}^{j_n} \nabla_{20}^{k_1} \cdots \nabla_{20}^{k_p} \frac{K_1(x_{10} + x_{20} +
    x_{12})}{(x_{10}\, x_{20}\, x_{12})}\,,\nn
&& \mathcal{F} \left( \int\frac{d^3 k_1}{(2\pi)^3}d^3 k_2 d^3 k_3\,
\delta (\vec k_2-\vec k_1+\vec q_1) \, \delta ( \vec k_3- \vec k_1- \vec
q_2) \, k_1^{i_1}\cdots k_1^{i_m} k_2^{j_1}\cdots
  k_2^{j_n} k_3^{k_1} \cdots k_3^{k_p} \, D_3 \right) \nn
&=& (-1)^{m+p} \frac{M_\pi^2 (i M_\pi)^{m+n+p}}{32\pi^4} \nabla_{12}^{i_1}\cdots\nabla_{12}^{i_m} \nabla_{10}^{j_1}\cdots
  \nabla_{10}^{j_n} \nabla_{20}^{k_1} \cdots \nabla_{20}^{k_p} \, 
 \frac{x_{10} + x_{20}  +x_{12}}{x_{10} \, x_{20} \, x_{12}} K_0(x_{10} +
x_{20} + x_{12})\,.
\eeqa
%{\bf I think, we should proceed in the following way. We should first give
%  expressions for the currents in terms of non-simplified loop integrals
%  (i.e. integrand expressed in terms of $\omega_i$'s). This should be done in
%  the main text at the very beginning and would provide a useful interface for
%  the reader to be able to follow the details of the calculations. Then, using
%  the expressions in this appendix for Fourier transformations of loop
%  integrals in terms of $\omega$'s, it would be straightforward for the reader
%  to verify the results in coordinate space given in the main text. } 

%\setlength{\bibsep}{0.2em}
%\bibliographystyle{h-physrev3}
%\bibliography{/home/epelbaum/refs_h-elsevier3}

\begin{thebibliography}{10}

%\cite{Bedaque:2002mn}
\bibitem{Bedaque:2002mn}
P.~F.~Bedaque and U.~van Kolck,
%``Effective field theory for few-nucleon systems,''
Ann.\ Rev.\ Nucl.\ Part.\ Sci.\  {\bf 52}, 339 (2002).
%[arXiv:nucl-th/0203055].
%%CITATION = NUCL-TH 0203055;%%

%\cite{Epelbaum:2005pn}
\bibitem{Epelbaum:2005pn}
  E.~Epelbaum,
  %``Few-nucleon forces and systems in chiral effective field theory,''
  Prog.\ Part.\ Nucl.\ Phys.\  {\bf 57}, 654 (2006).
%  [arXiv:nucl-th/0509032].
  %%CITATION = NUCL-TH 0509032;%%

%\cite{Epelbaum:2008ga}
\bibitem{Epelbaum:2008ga}
  E.~Epelbaum, H.~W.~Hammer and U.-G.~Mei{\ss}ner,
  %``Modern Theory of Nuclear Forces,''
  arXiv:0811.1338 [nucl-th], to appear in Rev. Mod. Phys..
  %%CITATION = ARXIV:0811.1338;%%

%\cite{Park:1993jf}
\bibitem{Park:1993jf}
  T.~S.~Park, D.~P.~Min and M.~Rho,
  %``Chiral dynamics and heavy fermion formalism in nuclei. 1. Exchange axial
  %currents,''
  Phys.\ Rept.\  {\bf 233}, 341 (1993)
  [arXiv:hep-ph/9301295].
  %%CITATION = PRPLC,233,341;%%

%\cite{Park:1995pn}
\bibitem{Park:1995pn}
  T.~S.~Park, D.~P.~Min and M.~Rho,
  %``Chiral Lagrangian approach to exchange vector currents in nuclei,''
  Nucl.\ Phys.\  A {\bf 596}, 515 (1996)
  [arXiv:nucl-th/9505017].
  %%CITATION = NUPHA,A596,515;%%

%\cite{Song:2007bj}
\bibitem{Song:2007bj}
  Y.~H.~Song, R.~Lazauskas, T.~S.~Park and D.~P.~Min,
  %``Effective field theory approach for the M1 properties of A=2 and 3
  %nuclei,''
  Phys.\ Lett.\  B {\bf 656}, 174 (2007)
  [arXiv:0705.2657 [nucl-th]].
  %%CITATION = PHLTA,B656,174;%%

%\cite{Song:2008zf}
\bibitem{Song:2008zf}
  Y.-H.~Song, R.~Lazauskas and T.-S.~Park,
  % Up-to N$^3$LO heavy-baryon chiral perturbation theory
  %                calculation for the M1 properties of three-nucleon
  %                systems
  arXiv:0812.3834 [nucl-th].
  %% CITATION = 0812.3834;%%

%\cite{Lazauskas:2009nw}
\bibitem{Lazauskas:2009nw}
  R.~Lazauskas, Y.~H.~Song and T.~S.~Park,
  %``Heavy-baryon chiral perturbation theory approach to thermal neutron capture
  %on ${}^{3}{He}$,''
  arXiv:0905.3119 [nucl-th].
  %%CITATION = ARXIV:0905.3119;%%

%\cite{Park:1999sz}
\bibitem{Park:1999sz}
  T.~S.~Park, K.~Kubodera, D.~P.~Min and M.~Rho,
  %``Effective field theory approach to n(pol.) + p(pol.) --> d + gamma at
  %threshold,''
  Phys.\ Lett.\  B {\bf 472}, 232 (2000)
  [arXiv:nucl-th/9906005].
  %%CITATION = PHLTA,B472,232;%%

%\cite{Park:1998wq}
\bibitem{Park:1998wq}
  T.~S.~Park, K.~Kubodera, D.~P.~Min and M.~Rho,
  %``The solar proton burning process revisited in chiral perturbation
  %theory,''
  Astrophys.\ J.\  {\bf 507}, 443 (1998)
  [arXiv:astro-ph/9804144].
  %%CITATION = ASJOA,507,443;%%

%\cite{Park:2001jn}
\bibitem{Park:2001jn}
  T.~S.~Park {\it et al.},
  %``The solar hep process in effective field theory,''
  arXiv:nucl-th/0107012.
  %%CITATION = NUCL-TH/0107012;%%

%\cite{Ando:2001es}
\bibitem{Ando:2001es}
  S.~Ando, T.~S.~Park, K.~Kubodera and F.~Myhrer,
  %``The mu- d capture rate in effective field theory,''
  Phys.\ Lett.\  B {\bf 533}, 25 (2002)
  [arXiv:nucl-th/0109053].
  %%CITATION = PHLTA,B533,25;%%

%\cite{Park:2002yp}
\bibitem{Park:2002yp}
  T.~S.~Park {\it et al.},
  %``Parameter-free effective field theory calculation for the solar proton
  %fusion and hep processes,''
  Phys.\ Rev.\  C {\bf 67}, 055206 (2003)
  [arXiv:nucl-th/0208055].
  %%CITATION = PHRVA,C67,055206;%%

%\cite{Kubodera:2004zm}
\bibitem{Kubodera:2004zm}
  K.~Kubodera and T.~S.~Park,
  %``The solar He p process,''
  Ann.\ Rev.\ Nucl.\ Part.\ Sci.\  {\bf 54}, 19 (2004)
  [arXiv:nucl-th/0402008].
  %%CITATION = ARNUA,54,19;%%

%\cite{Phillips:1999am}
\bibitem{Phillips:1999am}
  D.~R.~Phillips and T.~D.~Cohen,
  %``Deuteron electromagnetic properties and the viability of effective  field
  %theory methods in the two-nucleon system,''
  Nucl.\ Phys.\  A {\bf 668}, 45 (2000)
  [arXiv:nucl-th/9906091].
  %%CITATION = NUPHA,A668,45;%%

%\cite{Walzl:2001vb}
\bibitem{Walzl:2001vb}
  M.~Walzl and U.-G.~Mei{\ss}ner,
  %``Elastic electron deuteron scattering in chiral effective field theory,''
  Phys.\ Lett.\  B {\bf 513}, 37 (2001)
  [arXiv:nucl-th/0103020].
  %%CITATION = PHLTA,B513,37;%%

%\cite{Phillips:2003jz}
\bibitem{Phillips:2003jz}
  D.~R.~Phillips,
  %``Higher-order calculations of electron deuteron scattering in nuclear
  %effective theory,''
  Phys.\ Lett.\  B {\bf 567}, 12 (2003)
  [arXiv:nucl-th/0304046].
  %%CITATION = PHLTA,B567,12;%%

%\cite{Phillips:2006im}
\bibitem{Phillips:2006im}
  D.~R.~Phillips,
  %``Chiral effective theory predictions for deuteron form factor ratios at low
  %Q**2,''
  J.\ Phys.\ G {\bf 34}, 365 (2007)
  [arXiv:nucl-th/0608036].
  %%CITATION = JPHGB,G34,365;%%

%\cite{Valderrama:2007ja}
\bibitem{Valderrama:2007ja}
  M.~P.~Valderrama, A.~Nogga, E.~Ruiz Arriola and D.~R.~Phillips,
  %``Deuteron form factors in chiral effective theory: regulator-independent
  %results and the role of two-pion exchange,''
  Eur.\ Phys.\ J.\  A {\bf 36}, 315 (2008)
  [arXiv:0711.4785 [nucl-th]].
  %%CITATION = EPHJA,A36,315;%%

%\cite{Beane:2004ra}
\bibitem{Beane:2004ra}
  S.~R.~Beane, M.~Malheiro, J.~A.~McGovern, D.~R.~Phillips and U.~van Kolck,
  %``Compton scattering on the proton, neutron, and deuteron in chiral
  %perturbation theory to O(Q^4),''
  Nucl.\ Phys.\  A {\bf 747}, 311 (2005)
  [arXiv:nucl-th/0403088].
  %%CITATION = NUPHA,A747,311;%%

%\cite{Choudhury:2004yz}
\bibitem{Choudhury:2004yz}
  D.~Choudhury and D.~R.~Phillips,
  %``Predictions for polarized beam / vector-polarized target observables in
  %elastic Compton scattering on the deuteron,''
  Phys.\ Rev.\  C {\bf 71}, 044002 (2005)
  [arXiv:nucl-th/0411001].
  %%CITATION = PHRVA,C71,044002;%%

%\cite{Shukla:2008zc}
\bibitem{Shukla:2008zc}
  D.~Shukla, A.~Nogga and D.~R.~Phillips,
  %``Analyzing the Effects of Neutron Polarizabilities in Elastic Compton
  %Scattering off ${}^3He$,''
  Nucl.\ Phys.\  A {\bf 819}, 98 (2009)
  [arXiv:0812.0138 [nucl-th]].
  %%CITATION = NUPHA,A819,98;%%

%\cite{Beane:1997iv}
\bibitem{Beane:1997iv}
  S.~R.~Beane, V.~Bernard, T.~S.~H.~Lee, U.-G.~H.~Mei{\ss}ner and U.~van Kolck,
  %``Neutral pion photoproduction on deuterium in baryon chiral perturbation
  %theory to order q**4,''
  Nucl.\ Phys.\  A {\bf 618}, 381 (1997)
  [arXiv:hep-ph/9702226].
  %%CITATION = NUPHA,A618,381;%%

%\cite{Bernard:1999ff}
\bibitem{Bernard:1999ff}
  V.~Bernard, H.~Krebs and U.-G.~Mei{\ss}ner,
  %``On neutral pion electroproduction off deuterium,''
  Phys.\ Rev.\  C {\bf 61}, 058201 (2000)
  [arXiv:nucl-th/9912033].
  %%CITATION = PHRVA,C61,058201;%%

%\cite{Krebs:2002qr}
\bibitem{Krebs:2002qr}
  H.~Krebs, V.~Bernard and U.-G.~Mei{\ss}ner,
  %``Near threshold neutral pion electroproduction on deuterium in chiral
  %perturbation theory,''
  Nucl.\ Phys.\  A {\bf 713}, 405 (2003)
  [arXiv:nucl-th/0207072].
  %%CITATION = NUPHA,A713,405;%%

%\cite{Krebs:2004ir}
\bibitem{Krebs:2004ir}
  H.~Krebs, V.~Bernard and U.-G.~Mei{\ss}ner,
  %``Improved analysis of neutral pion electroproduction off deuterium in
  %chiral perturbation theory,''
  Eur.\ Phys.\ J.\  A {\bf 22}, 503 (2004)
  [arXiv:nucl-th/0405006].
  %%CITATION = EPHJA,A22,503;%%

%\cite{Gardestig:2005pp}
\bibitem{Gardestig:2005pp}
  A.~Gardestig and D.~R.~Phillips,
  %``Using chiral perturbation theory to extract the neutron neutron  scattering
  %length from pi- d --> n n gamma,''
  Phys.\ Rev.\  C {\bf 73}, 014002 (2006)
  [arXiv:nucl-th/0501049].
  %%CITATION = PHRVA,C73,014002;%%

%\cite{Lensky:2005hb}
\bibitem{Lensky:2005hb}
  V.~Lensky, V.~Baru, J.~Haidenbauer, C.~Hanhart, A.~E.~Kudryavtsev and U.-G.~Mei{\ss}ner,
  %``Precision calculation of gamma d --> pi+ n n within chiral perturbation
  %theory,''
  Eur.\ Phys.\ J.\  A {\bf 26}, 107 (2005)
  [arXiv:nucl-th/0505039].
  %%CITATION = EPHJA,A26,107;%%

%\cite{Golak:2005iy}
\bibitem{Golak:2005iy}
  J.~Golak, R.~Skibinski, H.~Witala, W.~Gl\"ockle, A.~Nogga and H.~Kamada,
  %``Electron and Photon Scattering on Three-Nucleon Bound States,''
  Phys.\ Rept.\  {\bf 415}, 89 (2005)
  [arXiv:nucl-th/0505072].
  %%CITATION = PRPLC,415,89;%%

%\cite{Epelbaum:2004fk}
\bibitem{Epelbaum:2004fk}
  E.~Epelbaum, W.~Gl\"ockle and U.-G.~Mei{\ss}ner,
  %``The two-nucleon system at next-to-next-to-next-to-leading order,''
  Nucl.\ Phys.\  A {\bf 747}, 362 (2005)
  [arXiv:nucl-th/0405048].
  %%CITATION = NUPHA,A747,362;%%

%\cite{Entem:2003ft}
\bibitem{Entem:2003ft}
  D.~R.~Entem and R.~Machleidt,
  %``Accurate Charge-Dependent Nucleon-Nucleon Potential at Fourth Order of
  %Chiral Perturbation Theory,''
  Phys.\ Rev.\  C {\bf 68}, 041001 (2003)
  [arXiv:nucl-th/0304018].
  %%CITATION = PHRVA,C68,041001;%%

%\cite{Epelbaum:2002vt}
\bibitem{Epelbaum:2002vt}
  E.~Epelbaum, A.~Nogga, W.~Gl\"ockle, H.~Kamada, U.-G.~Mei{\ss}ner and H.~Witala,
  %``Three-nucleon forces from chiral effective field theory,''
  Phys.\ Rev.\  C {\bf 66}, 064001 (2002)
  [arXiv:nucl-th/0208023].
  %%CITATION = PHRVA,C66,064001;%%

%\cite{Pastore:2008ui}
\bibitem{Pastore:2008ui}
  S.~Pastore, R.~Schiavilla and J.~L.~Goity,
  %``Electromagnetic two-body currents of one- and two-pion range,''
  Phys.\ Rev.\  C {\bf 78}, 064002 (2008)
  [arXiv:0810.1941 [nucl-th]].
  %%CITATION = PHRVA,C78,064002;%%

%\cite{Pastore:2009is}
\bibitem{Pastore:2009is}
  S.~Pastore, L.~Girlanda, R.~Schiavilla, M.~Viviani and R.~B.~Wiringa,
  %``Electromagnetic Currents and Magnetic Moments in $\chi$EFT,''
  arXiv:0906.1800 [nucl-th].
  %%CITATION = ARXIV:0906.1800;%%

%\cite{Epelbaum:1998ka}
\bibitem{Epelbaum:1998ka}
  E.~Epelbaum, W.~Gl\"ockle and U.-G.~Mei{\ss}ner,
  %``Nuclear forces from chiral Lagrangians using the method of unitary
  %transformation. I: Formalism,''
  Nucl.\ Phys.\  A {\bf 637}, 107 (1998)
  [arXiv:nucl-th/9801064].
  %%CITATION = NUPHA,A637,107;%%

%\cite{Epelbaum:1999dj}
\bibitem{Epelbaum:1999dj}
  E.~Epelbaum, W.~Gl\"ockle and U.-G.~Mei{\ss}ner,
  %``Nuclear forces from chiral Lagrangians using the method of unitary
  %transformation. II: The two-nucleon system,''
  Nucl.\ Phys.\  A {\bf 671}, 295 (2000)
  [arXiv:nucl-th/9910064].
  %%CITATION = NUPHA,A671,295;%%

%\cite{Epelbaum:2002gb}
\bibitem{Epelbaum:2002gb}
  E.~Epelbaum, U.-G.~Mei{\ss}ner and W.~Gl\"ockle,
  %``Nuclear forces in the chiral limit,''
  Nucl.\ Phys.\  A {\bf 714}, 535 (2003)
  [arXiv:nucl-th/0207089].
  %%CITATION = NUPHA,A714,535;%%

%\cite{Epelbaum:2006eu}
\bibitem{Epelbaum:2006eu}
  E.~Epelbaum,
  %``Four-Nucleon Force In Chiral Effective Field Theory,''
  Phys.\ Lett.\  B {\bf 639}, 456 (2006).
  %%CITATION = PHLTA,B639,456;%%

%\cite{Epelbaum:2007us}
\bibitem{Epelbaum:2007us}
  E.~Epelbaum,
  %``Four-nucleon force using the method of unitary transformation,''
  Eur.\ Phys.\ J.\  A {\bf 34}, 197 (2007)
  [arXiv:0710.4250 [nucl-th]].
  %%CITATION = EPHJA,A34,197;%%

\bibitem{Fukuda:1954aa}
N.~Fukuda, K.~Sawada, and M.~Taketani, Prog.\ Theor.\ Phys.\ {\bf 12}, 
156 (1954).

\bibitem{Okubo:1954aa}
S.~Okubo, Prog.\ Theor.\ Phys.\ {\bf 12}, 603 (1954).

%\cite{Krebs:2004st}
\bibitem{Krebs:2004st}
  H.~Krebs, V.~Bernard and U.-G.~Mei{\ss}ner,
  %``Orthonormalization procedure for chiral effective nuclear field theory,''
  Annals Phys.\  {\bf 316}, 160 (2005)
  [arXiv:nucl-th/0407078].
  %%CITATION = APNYA,316,160;%%

%\cite{Gari:1976kj}
\bibitem{Gari:1976kj}
  M.~Gari and H.~Hyuga,
  %``Mesonic Degrees Of Freedom In Nuclei And The Definition Of Meson Exchange
  %Currents,''
  Z.\ Phys.\  A {\bf 277}, 291 (1976).
  %%CITATION = ZEPYA,A277,291;%%

%\cite{Hyuga:1977cj}
\bibitem{Hyuga:1977cj}
  H.~Hyuga and H.~Ohtsubo,
  %``Breakdown Of The Siegert Theorem And The Many Body Charge Density
  %Operators,''
  Nucl.\ Phys.\  A {\bf 294}, 348 (1978).
  %%CITATION = NUPHA,A294,348;%%

%\cite{Eden:1995rf}
\bibitem{Eden:1995rf}
  J.~A.~Eden and M.~F.~Gari,
  %``A Consistent meson field theoretical description of p p Bremsstrahlung,''
  Phys.\ Rev.\  C {\bf 53}, 1102 (1996)
  [arXiv:nucl-th/9506001].
  %%CITATION = PHRVA,C53,1102;%%


%\cite{Koelling:future}
\bibitem{Koelling:future}
  S.~K\"olling, E.~Epelbaum, H.~Krebs, U.-G.~Mei{\ss}ner, \emph{in preparation}.

%\cite{Gerstein:1971fm}
\bibitem{Gerstein:1971fm}
  I.~S.~Gerstein, R.~Jackiw, S.~Weinberg and B.~W.~Lee,
  %``Chiral loops,''
  Phys.\ Rev.\  D {\bf 3}, 2486 (1971).
  %%CITATION = PHRVA,D3,2486;%%

%\cite{Davydychev:1991va}
\bibitem{Davydychev:1991va}
  A.~I.~Davydychev,
  %``A Simple formula for reducing Feynman diagrams to scalar integrals,''
  Phys.\ Lett.\  B {\bf 263}, 107 (1991).
  %%CITATION = PHLTA,B263,107;%%

%\cite{Gasser:1984ux}
\bibitem{Gasser:1984ux}
  J.~Gasser and H.~Leutwyler,
  %``Low-Energy Expansion Of Meson Form-Factors,''
  Nucl.\ Phys.\  B {\bf 250}, 517 (1985).
  %%CITATION = NUPHA,B250,517;%%

%\cite{Ecker:1995rk}
\bibitem{Ecker:1995rk}
  G.~Ecker and M.~Moj\v zi\v s,
  %``Low-energy expansion of the pion - nucleon Lagrangian,''
  Phys.\ Lett.\  B {\bf 365}, 312 (1996)
  [arXiv:hep-ph/9508204].
  %%CITATION = PHLTA,B365,312;%%

%\cite{Fettes:1998ud}
\bibitem{Fettes:1998ud}
  N.~Fettes, U.-G.~Mei{\ss}ner and S.~Steininger,
  %``Pion nucleon scattering in chiral perturbation theory.  I:
  %Isospin-symmetric case,''
  Nucl.\ Phys.\  A {\bf 640}, 199 (1998)
  [arXiv:hep-ph/9803266].
  %%CITATION = NUPHA,A640,199;%%

%\cite{Gasser:2002am}
\bibitem{Gasser:2002am}
  J.~Gasser, M.~A.~Ivanov, E.~Lipartia, M.~Moj\v zi\v s and A.~Rusetsky,
  %``Ground-state energy of pionic hydrogen to one loop,''
  Eur.\ Phys.\ J.\  C {\bf 26}, 13 (2002)
  [arXiv:hep-ph/0206068].
  %%CITATION = EPHJA,C26,13;%%

%\cite{Kaiser:1997mw}
\bibitem{Kaiser:1997mw}
  N.~Kaiser, R.~Brockmann and W.~Weise,
  %``Peripheral nucleon nucleon phase shifts and chiral symmetry,''
  Nucl.\ Phys.\  A {\bf 625}, 758 (1997)
  [arXiv:nucl-th/9706045].
  %%CITATION = NUPHA,A625,758;%%

\bibitem{Hemmert:1997ye}
T.~R. Hemmert, B.~R. Holstein, and J.~Kambor,
\newblock J. Phys. G {\bf 24}, 1831 (1998), hep-ph/9712496.
%%CITATION = HEP-PH 9712496;%%

%\cite{Kaiser:1998wa}
\bibitem{Kaiser:1998wa}
  N.~Kaiser, S.~Gerstendorfer and W.~Weise,
  %``Peripheral N N scattering: Role of Delta excitation, correlated  two-pion
  %and vector meson exchange,''
  Nucl.\ Phys.\  A {\bf 637}, 395 (1998)
  [arXiv:nucl-th/9802071].
  %%CITATION = NUPHA,A637,395;%%

\bibitem{Ordonez:1995rz}
C.~Ord\'o\~nez, L.~Ray, and U.~van Kolck,
\newblock Phys. Rev. C {\bf 53}, 2086 (1996), hep-ph/9511380.
%%CITATION = HEP-PH 9511380;%%

%\cite{Krebs:2007rh}
\bibitem{Krebs:2007rh}
  H.~Krebs, E.~Epelbaum and U.-G.~Mei{\ss}ner,
  %``Nuclear forces with Delta-excitations up to next-to-next-to-leading order
  %I: peripheral nucleon-nucleon waves,''
  Eur.\ Phys.\ J.\  A {\bf 32}, 127 (2007)
  [arXiv:nucl-th/0703087].
  %%CITATION = EPHJA,A32,127;%%

%\cite{Epelbaum:2007sq}
\bibitem{Epelbaum:2007sq}
  E.~Epelbaum, H.~Krebs and U.-G.~Mei{\ss}ner,
  %``Delta-excitations and the three-nucleon force,''
  Nucl.\ Phys.\  A {\bf 806}, 65 (2008)
  [arXiv:0712.1969 [nucl-th]].
  %%CITATION = NUPHA,A806,65;%%

%\cite{Epelbaum:2008td}
\bibitem{Epelbaum:2008td}
  E.~Epelbaum, H.~Krebs and U.-G.~Mei{\ss}ner,
  %``Isospin-breaking two-nucleon force with explicit Delta-excitations,''
  arXiv:0801.1299 [nucl-th].
  %%CITATION = ARXIV:0801.1299;%%

%\cite{Pandharipande:2005sx}
\bibitem{Pandharipande:2005sx}
  V.~R.~Pandharipande, D.~R.~Phillips and U.~van Kolck,
  %Delta effects in pion nucleon scattering and the strength
  %                of the  two-pion-exchange three-nucleon interaction
  \newblock Phys. Rev. C {\bf 71}, 064002 (2005), hep-ph/0501061.
  %%CITATION = NUCL-TH/0501061;%%

%\cite{Foldy:1979ie}
\bibitem{Foldy:1979ie}
  L.~L.~Foldy and J.~A.~Lock,
  %``Symmetries, Conservation Principles, And The Phenomenology Of Meson
  %Exchange Currents,''
%\href{/spires/find/hep/www?irn=641332}{SPIRES entry} 
in M.~Rho, D.~Wilkinson, {\it Mesons In Nuclei}, Vol.II, Amsterdam 1979, 465-493.

\end{thebibliography}
%\end{document}

\end{document}